\documentclass[reqno,11pt,nosumlimits]{article}
\usepackage{amssymb,amsfonts,amscd}
\usepackage[mathscr]{eucal}
\usepackage{amsmath}
\usepackage{graphics}
\usepackage{color}

\usepackage{graphicx}
\usepackage{ulem}

\allowdisplaybreaks[1]
\numberwithin{equation}{section}

\renewcommand{\em}{\it}

\renewcommand{\l}{\lambda}
\renewcommand{\th}{\theta}
\renewcommand{\r}{\rho}
\renewcommand{\a}{\alpha}

\newtheorem{theorem}{Theorem}[section]
\newtheorem{proposition}[theorem]{Proposition} 

\newtheorem{remark}{Remark}[section]

\newtheorem{convention}{Convention}
\newtheorem{assumption}{Assumption}

\newtheorem{conjecture}{Conjecture}

\newtheorem{question}{Question}

\newcommand{\vf}{\varphi}

\newcommand{\cl}{{\mathrm h}}

\newcommand{\orth}{\perp}

\newcommand{\id}{{ \rm id}}

\newcommand{\bN}{{\mathbb N}}
\newcommand{\bR}{{\mathbb R}}
\newcommand{\bC}{{\mathbb C}}
\newcommand{\bZ}{{\mathbb Z}}

\newcommand{\cA}{{\mathcal A}}
\newcommand{\ck}{\mathfrak k}

\newcommand{\cS}{{\mathscr S}}
\newcommand{\cB}{{\mathcal B}}
\newcommand{\cO}{{\mathcal O}}
\newcommand{\cF}{{\mathcal F}}
\newcommand{\cR}{{\mathcal R}}
\newcommand{\cD}{{\mathcal D}}
\newcommand{\cN}{{\mathcal N}}

\newcommand{\ct}{{\mathfrak t}}

\newcommand{\cG}{{\mathfrak g}}

\newcommand{\cH}{{\mathcal H}}

\newcommand{\cC}{{\mathcal C}}
\newcommand{\cE}{{\mathcal E}}

\newcommand{\eps}{\epsilon}

\newcommand{\cg}{c_{\cG}}
\newcommand{\cW}{{\mathcal W}}

\newcommand{\faces}{\cF}

\DeclareMathOperator{\Image}{Image}
\DeclareMathOperator{\arc}{arc}
 \DeclareMathOperator{\Tr}{Tr}
\DeclareMathOperator{\GL}{GL}
 \DeclareMathOperator{\Mat}{Mat}
\DeclareMathOperator{\supp}{supp}
 \DeclareMathOperator{\Det}{Det}

 \DeclareMathOperator{\End}{End}
\DeclareMathOperator{\Hom}{Hom}
\DeclareMathOperator{\Ad}{Ad}
\DeclareMathOperator{\ad}{ad}

\DeclareMathOperator{\Aff}{Aff}

 \DeclareMathOperator{\WLO}{WLO}
\DeclareMathOperator{\Hol}{Hol}

\DeclareMathOperator{\wind}{wind}

\DeclareMathOperator{\sgn}{sgn}

\DeclareMathOperator{\Ker}{ker}

\DeclareMathOperator{\gleam}{gleam}

\DeclareMathOperator{\aff}{aff}

\begin{document}

\title{Infinite dimensional analysis and the Chern-Simons path integral}

\maketitle

\begin{center} \Large
Atle Hahn
\end{center}

\begin{center}
\it   \large  Grupo de  F{\'i}sica Matem{\'a}tica da Universidade de Lisboa \\
Av. Prof. Gama Pinto, 2\\
PT-1649-003 Lisboa, Portugal\\
Email: atle.hahn@gmx.de
  \end{center}

\begin{abstract}

Using the framework of White Noise Analysis we give
a rigorous implementation of the gauge fixed
Chern-Simons path integral associated to an arbitrary simple simply-connected compact  structure group $G$
and a simple class of (ribbon) links $L$ in  the base manifold $M=S^2 \times S^1$.\par

\end{abstract}

\medskip


\medskip


\medskip

\section{Introduction}
\label{sec1}

In the  1980s and the beginning of the 1990s
Jones, Witten,  Turaev,  Reshetikhin,  Kontsevich,
and others revolutionized Knot Theory and  created
a whole new area which is now called ``Quantum Topology''.
Quantum Topology is both a deep and  a beautiful  theory, beautiful in the sense that
 it naturally connects a large number of branches of Mathematics and Physics\footnote{this is reflected also in the present paper:   Quantum Gauge Field Theory plays the main role in  Sec. \ref{subsec2.1b} and Sec. \ref{subsec2.2},
 Riemannian Geometry/Global Analysis  in Sec. \ref{subsec2.5},  Infinite Dimensional Analysis in Sec. \ref{sec3}, and
 Algebra and low-dimensional   Topology  in  Appendix \ref{appA}}  like Algebra (Lie algebras, affine Lie algebras, quantum groups, ..., and the corresponding
representation theories), low-dimensional Topology \& Knot Theory,
Riemannian Geometry \& Global Analysis,  Infinite Dimensional Analysis, and
 Quantum Field Theory (Gauge Field Theory, Conformal Field Theory, Quantum Gravity,  String Theory).\par

The first major step towards Quantum Topology was
the discovery of the Jones polynomial and its generalizations (in particular, the HOMFLY and the Kauffman polynomials) in 1984 and 1985. In 1988,  Witten  demonstrated in a celebrated paper \cite{Wi} that
the heuristic Feynman path integral associated
to  a certain 3-dimensional gauge field theory
can be used to give a very elegant and intrinsically  3-dimensional
 ``definition'' of the Jones polynomial and the other knot
polynomials mentioned above.
The aforementioned gauge theory is the so-called
(pure) ``Chern-Simons model'' which is specified by a triple $(M,G,k)$
where  $M$ is an oriented connected 3-dimensional manifold (usually compact),
$G$ is a semi-simple Lie group (often compact and simply-connected), and $k \in \bN$
is a fixed parameter (``the level'').
In the very important special case $G=SU(N)$, $N \ge 2$,
 the action function $S_{CS}$ of the Chern-Simons model is given explicitly by
\begin{equation} \label{eq_intro}
S_{CS}(A) = k \int_M \Tr(A\wedge dA + \tfrac{2}{3} A \wedge A \wedge A), \quad \quad A  \in \cA
\end{equation}
where $\cA$ is the space of all $su(N)$-valued 1-forms $A$ on $M$,
$\wedge$ the wedge product  associated to the multiplication of $\Mat(N,\bC)$,
   and $\Tr:\Mat(N,\bC) \to \bC$ the suitably normalized trace. \par
The  Chern-Simons  ``path space  measure'' is the informal complex measure
$\exp(i S_{CS}(A)) DA$ where $DA$ is the (ill-defined) Lebesgue measure on the infinite dimensional space $\cA$.
The  Chern-Simons  path integral (functional) $\int \cdots \exp(i S_{CS}(A)) DA$ not only
 gives a unifying framework for the aforementioned knot polynomials\footnote{\label{ft_2}
 the polynomial invariants mentioned above
 are given by $\WLO(L)$ as defined in Eq. \eqref{eq_WLO_orig} below.
In the special case $M = S^3$, $G = SU(2)$, and where each representation
$\rho_i$ appearing in in Eq. \eqref{eq_WLO_orig} is the fundamental representation of $SU(2)$
$\WLO(L)$ is given by an explicit expression
involving the Jones polynomial of $L$. }
  but also leads naturally to
     a generalization of these invariants to all closed 3-manifolds, the ``Jones-Witten invariants''.\par

Using several heuristic arguments,
some of them from Conformal Field Theory, Witten was able to evaluate
the Jones-Witten invariants explicitly.
Two years later  Reshetikhin
and Turaev  finally found a rigorous (and equivalent) version of the Jones-Witten invariants,
the so-called  ``Reshetikhin-Turaev invariants'',  \cite{ReTu1,ReTu2}.
The approach in  \cite{ReTu1,ReTu2} is algebraic and very different from the path integral approach.
In particular, it is based on
the representation theory of quantum groups and ``surgery operations'' on the relevant 3-manifolds.
Turaev  \cite{Tu2} later found  an equivalent approach
called the ``shadow world'' approach, which is also based on quantum group representations
but uses certain finite ``state sums'' instead of the surgery operations.

\medskip

Many open questions in the field of Quantum Topology  are closely related
to the following  problem which is generally considered to be one of the
major open problems in the field (cf. \cite{Kup} und \cite{Sawin99}):

\begin{itemize}
\item[\bf (P1)] Find a rigorous realization\footnote{here the word ``realization'' is meant to imply that
the values of the rigorously defined Chern-Simons path integral expressions
coincide with the corresponding Reshetikhin-Turaev invariant}
 of the (original or gauge fixed) Chern-Simons path integral expressions
for all simply-connected compact Lie groups $G$ and  all (compact oriented 3-dimensional) base manifolds $M$.
\end{itemize}

Since (P1) seems to be a very hard problem it makes sense to restrict oneself
first to the following weakened version of (P1).

\begin{itemize}
\item[\bf (P1)'] Find a rigorous realization  of the (suitably gauge fixed) Chern-Simons path integral expressions
for all  compact and simply-connected Lie groups $G$  and some fixed base manifold $M$ like, e.g., $M=S^3$ or $M=S^2 \times S^1$.
\end{itemize}

 Note that the quadratic part of the Chern-Simons action function $S_{CS}$ is degenerate. Moreover, there is a cubic term, which is clearly not semi-bounded.
 Because of this we cannot hope to be able to transform
 the (original or gauge-fixed) Chern-Simons path space measure into a  bounded positive (heuristic) measure by
 applying a so-called ``Wick rotation''\footnote{this is different from the situation in many other bosonic QFTs
 where the path space measure can  be transformed
 into a  bounded positive (heuristic) measure by means of a  Wick rotation}.
 This means that the well-established theory of positive bounded measures on infinite dimensional
 topological  vector  spaces is not as useful as in standard Constructive QFT.
 Instead of working with bounded measures one can try to work with a suitable notion of distributions
 on infinite dimensional spaces as provided, e.g. by White Noise Analysis. \par

   And indeed, as we will  show in the present paper,  White Noise Analysis can  be used
 to make progress towards the solution of Problem (P1)'.\par
More precisely, for the special manifold  $M= S^2 \times S^1$
we will  give a rigorous realization of Witten's path integrals
 (after a suitable gauge fixing has been applied).
We expect (cf. Conjecture \ref{conj3} below)  that at least
for a simple type of (ribbon) links $L$
the aforementioned rigorous realization  reproduces  Turaev's shadow invariant. \par

Our paper is based on the ``torus gauge fixing'' approach
to Chern-Simons theory on base manifolds of the form $M=\Sigma \times S^1$
which was developed in \cite{BlTh1,BlTh2, BlTh3,Ha3b,Ha3c,Ha4,HaHa} (see also \cite{Ha7a}
for later developments). In \cite{Ha3b,Ha4,Ha6} we sketched how  a fully rigorous
realization of the torus gauge fixed Chern-Simons path integral
could be obtained  within the framework of White Noise Analysis (using ideas from \cite{LS,ASen,Ha2}).
 We remark here, however, that
several constructions in \cite{Ha3b,Ha4,Ha6} (and also in \cite{HaHa}) are unnecessarily complicated,
which is why in the present paper we will reconsider the issue
and make the following changes and improvements:

\begin{enumerate}

\item[1.] Instead of the heuristic formula Eq. (15) in \cite{Ha6}, which was used as the starting point for
  the treatment in \cite{Ha6}, we will use the simplified and more natural heuristic formula
 Eq. \eqref{eq2.48_Ha7a} below (which was first derived in \cite{Ha7a}).
 In particular, the space $\hat{\cA}^{\orth}$ appearing in \cite{Ha6}
 is now replaced by the space $\Check{\cA}^{\orth}$. Moreover, the singular 1-forms $A^{\orth}_{sing}(\cl)$
 appearing in Eq. (15) in \cite{Ha6} are now absent.

\item[2.] We use an alternative rigorous implementation of the heuristic expression
$\Det(B):=\Det_{FP}(B)  \Check{Z}(B)$ in   Sec. \ref{subsec2.5}, cf. Eqs. \eqref{eq_Det_rig_defs} below.
The implementation in Sec. \ref{subsec2.5} is considerably more natural
than the  ``ad hoc ansatz''  in Eq. (13) in \cite{Ha6}  for $\Det(B)$.

\item[3.] The framing procedure is implemented in a different way.
      We no longer use a family of diffeomorphisms $(\phi_s)_{s > 0}$ of $M=\Sigma \times S^1$
      in order to deform the two Hida distributions appearing in \cite{Ha6}.
      Instead we will work with the undeformed Hida distributions but
      replace the (smeared) loops by (smeared) ribbons.
 This has two important advantages:
 Firstly, it makes the alternative, natural definition of $\Det(B)$ mentioned above possible.
 Secondly, it seems to eliminate the ``loop smearing dependence problem''
 which would probably have appeared by carrying out the original approach in \cite{Ha6}.

\end{enumerate}

We remark that in  \cite{Ha7a,Ha7b} we have developed an alternative ``simplicial'' approach for making sense of the RHS of Eq. \eqref{eq2.48_Ha7a},
see Sec. \ref{subsec4.2} below for a brief comparison of
the rigorous continuum approach of the present paper and the  simplicial approach in \cite{Ha7a,Ha7b}.

\smallskip

The present paper is organized as follows:\par

In Secs \ref{subsec2.1}--\ref{subsec2.2} we recall the relevant heuristic formulas for the
Wilson loop observables in Chern-Simons theory,
first the original formula  Eq. \eqref{eq_WLO_orig}
and later the modified formula which was obtained in \cite{Ha7a}
 by applying torus gauge fixing in the special case
 where $M=\Sigma \times S^1$, cf. Eq. \eqref{eq2.48_Ha7a}.
  In Sec. \ref{subsec2.4} we introduce ``infinitesimal ribbons''  and rewrite the heuristic formula Eq \eqref{eq2.48_Ha7a},
 obtaining  Eq. \eqref{eq2.48_ribbon}.
 In Sec. \ref{subsec2.5} we
 give the rigorous definition of the expression $\Det(B)=\Det_{FP}(B)  \Check{Z}(B)$
 mentioned above. In Sec. \ref{subsec2.6} we finally rewrite  the heuristic formula Eq. \eqref{eq2.48_ribbon}
  obtained in Sec. \ref{subsec2.4}  in a suitable  way (for the special case $\Sigma = S^2$) arriving at the heuristic formula Eq. \eqref{eq_def_FL}.\par

In Sec. \ref{sec3} we then explain how one can make rigorous sense of
the RHS of the aforementioned Eq.  \eqref{eq_def_FL}.
In Sec. \ref{sec4} we conclude the main part of this paper with a brief discussion of our results. \par

In Appendix \ref{appA} we briefly recall the definition of Turaev's shadow invariant in the
special case relevant for us, i.e.  $M=S^2 \times S^1$.
In Appendix \ref{appB} we  fill in some technical details which were omitted in Sec. \ref{sec3}.

\section{The heuristic Chern-Simons  path integral  in the torus gauge}
\label{sec2}

We fix a  simple  simply-connected compact Lie group
 $G$  and  a maximal torus $T$ of $G$.
By  $\cG$ and $\ct$ we  will denote the Lie algebras of $G$ and  $T$
and by $\langle \cdot , \cdot \rangle_{\cG}$ or simply by $\langle \cdot , \cdot \rangle$
 the unique $\Ad$-invariant scalar product
on $\cG$ satisfying the normalization condition $\langle \Check{\alpha} , \Check{\alpha} \rangle = 2$
for every short coroot $\Check{\alpha}$ w.r.t. $(\cG,\ct)$.

\smallskip

Moreover, we will fix  a compact oriented 3-manifold $M$
of the form $M = \Sigma \times S^1$ where $\Sigma$ is a compact oriented surface.
Finally, we fix an (ordered and oriented) ``link'' $L$ in $M$,
i.e. a finite tuple $L=(l_1, \ldots, l_m)$, $m \in \bN$, of pairwise non-intersecting
knots $l_i$ and we equip  each $l_i$ with a ``color'', i.e. a finite-dimensional complex representation $\rho_i$ of $G$.
Recall that a ``knot'' in $M$ is an embedding $l:S^1 \to M$.
Using the surjection $[0,1] \ni t \mapsto e^{2 \pi i t} \in S^1 \cong \{ z \in \bC \mid |z| =1 \}$
we can  consider each  knot  as a loop $l:[0,1] \to M$, $l(0) = l(1)$, in the obvious way.

\subsection{Basic  spaces}
\label{subsec2.1}

As in  \cite{Ha7a} we will use the following notation\footnote{Here  $\Omega^p(N,V)$ denotes the space of $V$-valued  $p$-forms
on a smooth manifold $N$}
\begin{subequations} \label{eq_basic_spaces_cont}
\begin{align}
\cB & = C^{\infty}(\Sigma,\ct)  \\
\cA & =  \Omega^1(M,\cG)\\
\cA_{\Sigma} & =  \Omega^1(\Sigma,\cG) \\
\cA_{\Sigma,\ct} & = \Omega^1(\Sigma,\ct), \quad  \cA_{\Sigma,\ck}  = \Omega^1(\Sigma,\ck) \\
\cA^{\orth} & =  \{ A \in \cA \mid A(\partial/\partial t) = 0\}\\
\label{eq_part_f}
\Check{\cA}^{\orth} & = \{ A^{\orth} \in \cA^{\orth} \mid \int A^{\orth}(t) dt \in \cA_{\Sigma,\ck} \} \\
\label{eq_part_g}   \cA^{\orth}_c & = \{ A^{\orth} \in \cA^{\orth} \mid \text{ $A^{\orth}$ is constant and
 $\cA_{\Sigma,\ct}$-valued}\}
 \end{align}
\end{subequations}
Here $\ck$ is the orthogonal complement of $\ct$ in $\cG$ w.r.t.
$\langle \cdot, \cdot \rangle$, $dt$ is the normalized translation-invariant volume form on $S^1$,
and $\partial/\partial t$ is the vector field on $M=\Sigma \times S^1$
obtained by ``lifting''  in the obvious way
the normalized translation-invariant vector field $\partial/\partial t$ on $S^1$.
Moreover,  in Eqs. \eqref{eq_part_f} and \eqref{eq_part_g}
 we used the ``obvious'' identification (cf. Sec. 2.3.1 in \cite{Ha7a})
\begin{equation}
\cA^{\orth}  \cong C^{\infty}(S^1,\cA_{\Sigma})
\end{equation}
where  $C^{\infty}(S^1,\cA_{\Sigma})$ is the space of maps
$f:S^1 \to \cA_{\Sigma}$ which are ``smooth'' in the sense that
$\Sigma \times S^1 \ni (\sigma,t) \mapsto (f(t))(X_{\sigma}) \in \cG$
is smooth for every smooth vector field $X$ on $\Sigma$.
It follows from the definitions above that
\begin{equation} \label{eq_cAorth_decomp}
\cA^{\orth} = \Check{\cA}^{\orth} \oplus  \cA^{\orth}_c
\end{equation}

\subsection{The heuristic Wilson loop observables}
\label{subsec2.1b}

The Chern-Simons action function $S_{CS}: \cA \to \bR$
associated to $M$, $G$, and the ``level''  $k \in \bZ \backslash \{0\}$ is
 given by\footnote{Eq. \eqref{eq2.2'} generalizes
Eq. \eqref{eq_intro} in Sec. \ref{sec1}. In Eq. \eqref{eq_intro}  the factor $- \pi$
is hidden  in the trace functional $\Tr:\Mat(N,\bC) \to \bC$.
Moreover, in Eq. \eqref{eq_intro} we have a factor $2/3$ instead of $1/3$ because
the wedge product $\wedge$ in Eq. \eqref{eq_intro}  differs from each of the two wedge
products in  Eq. \eqref{eq2.2'}}
 \begin{equation} \label{eq2.2'} S_{CS}(A) = - k \pi \int_M \langle A \wedge dA \rangle
   + \tfrac{1}{3} \langle A\wedge [A \wedge A]\rangle, \quad
A \in \cA \end{equation}
where $[\cdot \wedge \cdot]$  denotes the wedge  product associated to the
Lie bracket $[\cdot,\cdot] : \cG \times \cG \to \cG$
and where   $\langle \cdot \wedge  \cdot \rangle$
  denotes the wedge product  associated to the
 scalar product $\langle \cdot , \cdot \rangle : \cG \times \cG \to \bR$.\par

Recall  that the heuristic Wilson loop observable $\WLO(L)$ of a link $L=(l_1,l_2,\ldots,l_m)$ in $M$
with ``colors'' $(\rho_1,\rho_2,\ldots,\rho_m)$  is given by the informal ``path integral'' expression
\begin{equation} \label{eq_WLO_orig}
\WLO(L) := \int_{\cA} \prod_i  \Tr_{\rho_i}(\Hol_{l_i}(A)) \exp( i S_{CS}(A)) DA
\end{equation}
where $\Hol_l(A) \in G$ is the holonomy of $A \in \cA$ around the loop $l \in \{l_1, \ldots, l_m\}$
and where $DA$ is the (ill-defined) ``Lebesgue measure'' on the infinite-dimensional space $\cA$.
We will use the  following explicit formula for $\Hol_l(A)$
\begin{equation} \label{eq_Hol_heurist}
\Hol_l(A) = \lim_{n \to \infty} \prod_{j=1}^n \exp\bigl(\tfrac{1}{n}  A(l'(t))\bigr)_{| t=j/n}
\end{equation}
where $\exp:\cG \to G$ is the exponential map of $G$.

\subsection{The basic heuristic formula from \cite{Ha7a}}
\label{subsec2.2}

The starting point for the main part of \cite{Ha7a} was a second heuristic formula for $\WLO(L)$
which one obtains from Eq. \eqref{eq_WLO_orig} above after applying a suitable gauge fixing.

\smallskip

Let $\pi_{\Sigma}: \Sigma \times S^1 \to \Sigma$ be the canonical projection.
For each loop $l_i$ appearing in the link $L$ we set $l^i_{\Sigma}:= \pi_{\Sigma} \circ l_i$.
Moreover, we fix  $\sigma_0 \in \Sigma$ such that
$$\sigma_0 \notin \arc(L_{\Sigma}) := \bigcup_i \arc(l^i_{\Sigma})$$
By applying ``abstract torus gauge fixing'' (cf. Sec. 2.2.4 in  \cite{Ha7a})
and a suitable change of variable one can derive at a heuristic level  (cf. Eq. (2.53) in \cite{Ha7a})
\begin{multline}  \label{eq2.48_Ha7a} \WLO(L)
 \sim \sum_{y \in I}  \int_{\cA^{\orth}_c \times \cB} \biggl\{
 1_{\cB_{reg}}(B)  \Det_{FP}(B)\\
 \times   \biggl[ \int_{\Check{\cA}^{\orth}} \prod_i  \Tr_{\rho_i}\bigl(
 \Hol_{l_i}(\Check{A}^{\orth} + A^{\orth}_c, B)\bigr)
\exp(i  S_{CS}( \Check{A}^{\orth}, B)) D\Check{A}^{\orth} \biggr] \\
 \times \exp\bigl( - 2\pi i k  \langle y, B(\sigma_0) \rangle \bigr) \biggr\}
 \exp(i S_{CS}(A^{\orth}_c, B)) (DA^{\orth}_c \otimes DB)
\end{multline}
where ``$\sim$'' denotes equality up to a multiplicative ``constant''\footnote{``constant'' in the sense
 that  $C$ does not depend on $L$.
 By contrast, $C$ may  depend on $G$, $\Sigma$, and $k$.} $C$,
where $I:= \ker(\exp_{| \ct})  \subset \ct$, where $DB$ and $DA^{\orth}_c$ are the
 informal ``Lebesgue measures'' on the infinite-dimensional spaces $\cB$ and $\cA^{\orth}_c$,
 and where  have  set
 \begin{equation}
\cB_{reg}: = \{ B \in \cB \mid \forall \sigma \in \Sigma : B(\sigma) \in \ct_{reg} \} =  C^{\infty}(\Sigma,\ct_{reg})
\end{equation}
 with $\ct_{reg} := \exp^{-1}(T_{reg})$,
  $T_{reg}$ being the set of ``regular'' elements\footnote{i.e. the set of all $t \in T$ which are not contained
in a different maximal torus $T'$} of $T$.\par

Moreover, we have set for each $B \in \cB$, $A^{\orth} \in \cA^{\orth}$
\begin{align}
S_{CS}(A^{\orth},B) & := S_{CS}(A^{\orth} + B dt )\\
\label{eq_Hol_heurist_abbr} \Hol_{l}(A^{\orth},  B) &  := \Hol_{l}(A^{\orth}    + B dt)
\end{align}
Here $dt$ is the real-valued 1-form on  $M=\Sigma \times S^1$
obtained by  pulling back the 1-form $dt$ on $S^1$ by means of the canonical projection
$\pi_{S^1}: \Sigma \times S^1 \to S^1$.
Finally, $\Det_{FP}(B)$ is the informal expression   given by
\begin{equation} \label{eq_DetFP} \Det_{FP}(B) :=    \det\bigl(1_{\ck}-\exp(\ad(B))_{|\ck}\bigr)
\end{equation}
where $1_{\ck}-\exp(\ad(B))_{|\ck}$ is the linear operator on $C^{\infty}(\Sigma,\ck)$ given by
$$(1_{\ck}-\exp(\ad(B))_{|\ck} \cdot f)(\sigma) = (\id_{\ck}-\exp(\ad(B(\sigma)))_{|\ck}) \cdot f(\sigma)
\quad \quad \forall \sigma \in \Sigma, \quad \forall f \in   C^{\infty}(\Sigma,\ck)   $$
where on the RHS $\id_{\ck}$ is the identity on $\ck$.

\medskip

For the rest of this paper we will now
 fix an auxiliary Riemannian metric ${\mathbf g}={\mathbf g_{\Sigma}}$ on $\Sigma$.
 After doing so  we obtain  scalar products
   $\ll \cdot , \cdot \gg_{\cA_{\Sigma}}$ and
    $\ll \cdot , \cdot \gg_{\cA^{\orth}}$ on $\cA_{\Sigma}$ and
  $\cA^{\orth} \cong C^{\infty}(S^1, \cA_{\Sigma})$
 in a natural way. Moreover, we obtain a well-defined
Hodge star operator
 $\star: \cA_{\Sigma} \to \cA_{\Sigma}$ which induces an operator
 $\star: C^{\infty}(S^1, \cA_{\Sigma}) \to C^{\infty}(S^1, \cA_{\Sigma})$ in the obvious way, i.e.
 by $(\star A^{\orth})(t) = \star (A^{\orth}(t))$ for all $A^{\orth} \in \cA^{\orth}$ and $t \in S^1$.
 We  have the following explicit formula (cf. Eq. (2.48) in \cite{Ha7a})
 \begin{equation} \label{eq_SCS_expl0} S_{CS}(A^{\orth},B)  =  \pi k  \ll A^{\orth},
\star  \bigl(\tfrac{\partial}{\partial t} + \ad(B) \bigr) A^{\orth} \gg_{\cA^{\orth}}
 +  2 \pi k  \ll\star  A^{\orth},  dB \gg_{\cA^{\orth}}
\end{equation}
for all $B \in \cB$ and $A^{\orth} \in \cA^{\orth}$,
 which implies
 \begin{align} \label{eq_SCS_expl}
S_{CS}(\Check{A}^{\orth},B) & =  \pi k  \ll \Check{A}^{\orth},
\star  \bigl(\tfrac{\partial}{\partial t} + \ad(B) \bigr) \Check{A}^{\orth} \gg_{\cA^{\orth}} \\
\label{eq_SCS_expl2}
 S_{CS}(A^{\orth}_c,B) & =   2 \pi k  \ll\star  A^{\orth}_c,  dB \gg_{\cA^{\orth}}
\end{align}
for  $B \in \cB$, $\Check{A}^{\orth} \in \Check{\cA}^{\orth}$, and $A^{\orth}_c \in \cA^{\orth}_c$.

\medskip

The following informal definitions will be useful in  Sec. \ref{subsec2.4}  below: \par

For each $B \in \cB$ we set
\begin{align}  \label{eq_def_Z_B}
\Check{Z}(B) & := \int \exp(i  S_{CS}( \Check{A}^{\orth}, B)) D\Check{A}^{\orth},\\
 \label{eq_def_mu_B}
d\mu^{\orth}_B & := \tfrac{1}{\Check{Z}(B)} \exp(i  S_{CS}( \Check{A}^{\orth}, B)) D\Check{A}^{\orth}
\end{align}

Moreover, we will denote by $d_{\mathbf g}$ the distance function  on $\Sigma$
and by $d\mu_{\mathbf g}$ the volume measure on $\Sigma$ which
are associated to the Riemannian metric ${\mathbf g}$.

\subsection{Ribbon version  of Eq. \eqref{eq2.48_Ha7a}}
\label{subsec2.4}

It is well-known in the mathematics and physics literature on quantum 3-manifold
invariants that rather than  working with links one actually has to work
with framed links or, equivalently,
  with ribbon links (see below) if one wants to get meaningful results. \par

From the knot theory point of view
 the framed link picture and the ribbon link picture are equivalent.
 However,  the ribbon picture seems to be better suited for the
 study of the  Chern-Simons path integral in the torus gauge.

For every (closed) ribbon $R$ in $\Sigma \times S^1$, i.e. every smooth embedding
$R: S^1 \times [0,1] \to \Sigma \times S^1$,
we define
$$\Hol_{R}(A) :=  \lim_{n \to \infty} \prod_{j=1}^n \exp\bigl(\tfrac{1}{n} \int_{0}^1  A(R'_u(t)) du \bigr)_{| t=j/n}
 \in G$$
 where $R_u$, for  $u \in [0,1]$, is the loop in $\Sigma \times S^1$
given by $R_u(t) = R(t,u)$ for all $t \in S^1$.\smallskip

A ribbon link in $\Sigma \times S^1$ is
 a finite tuple of non-intersecting closed ribbons in in $\Sigma \times S^1$.
We will replace the link $L =(l_1,l_2, \ldots, l_m)$
by a ribbon link $L_{ribb} = (R_1,R_2, \ldots, R_m)$ where each $R_i$, $i \le m$,
is chosen such that  $l_i(t)= R_i(t,1/2)$ for all $t \in S^1$. \par

Moreover,  we ``scale'' each $R_i$, i.e. for each $s \in (0,1)$ we introduce
 the ribbon  $R^{(s)}_i$ by
$R^{(s)}_i(t,u) := R_i(t,s \cdot ( u  - 1/2) + 1/2)$ for all $t \in S^1$ and $u \in [0,1]$. \par

Then  we replace $\Hol_{l_i}(A)$ appearing in Eq. \eqref{eq2.48_Ha7a}
 (with $A = \Check{A}^{\orth} + A^{\orth}_c + Bdt$)
 by $\Hol_{R^{(s)}_i}(A)$. Moreover,  we include a $s \to 0$ limit.

\begin{convention}
We will usually write simply $L$ instead of $L_{ribb}$ when no confusion can arise.
\end{convention}

\begin{remark} The inclusion\footnote{if one does not include this limit $s  \to 0$ it may still be
possible to derive a result like Eq. \eqref{eq_theorem} below
for ribbon links $L$ fulfilling Assumption \ref{ass0} but most probably not for general ribbon links.}
of the limit $s \to 0$  above is the formal implementation of the intuitive idea
that our ribbons should  have ``infinitesimal width''.
\end{remark}

After these preparations  we arrive at the following ribbon analogue of Eq. \eqref{eq2.48_Ha7a} above
\begin{multline}  \label{eq2.48_ribbon} \WLO(L)
 \sim  \lim_{s\to 0} \sum_{y \in I}  \int_{\cA^{\orth}_c \times \cB} \biggl\{
 1_{\cB_{reg}}(B)  \Det_{FP}(B) \Check{Z}(B)\\
 \times   \biggl[ \int_{\Check{\cA}^{\orth}} \prod_i  \Tr_{\rho_i}\bigl(
 \Hol_{R^{(s)}_i}(\Check{A}^{\orth}, A^{\orth}_c, B)\bigr)  d\mu^{\orth}_B(\Check{A}^{\orth}) \biggr] \\
 \times \exp\bigl( - 2\pi i k  \langle y, B(\sigma_0) \rangle \bigr) \biggr\}
 \exp(i S_{CS}(A^{\orth}_c, B)) (DA^{\orth}_c \otimes DB)
\end{multline}
 where we have set (cf.  Sec. \ref{subsec2.2} above)
$$\Hol_{R^{(s)}_i}(\Check{A}^{\orth}, A^{\orth}_c, B) := \Hol_{R^{(s)}_i}(\Check{A}^{\orth} + A^{\orth}_c + Bdt)$$

In the following we set
$$R^i_{\Sigma} := \pi_{\Sigma} \circ R_i$$

From now on we will restrict ourselves to  ribbon links $L = (R_1,R_2, \ldots, R_m)$
fulfilling the following assumption.

\begin{assumption} \label{ass0} The maps $R^i_{\Sigma}$,  $i \le m$, neither intersect themselves nor each other.
More precisely: Each $R^i_{\Sigma}$,  $i \le m$, is an injection $S^1 \times  [0,1] \to \Sigma$
and we have $\Image(R^i_{\Sigma}) \cap \Image(R^j_{\Sigma}) = \emptyset$ if $i \neq j$.
\end{assumption}

\begin{remark} \label{rm_ass0}
It is interesting to consider also the weakened version of Assumption \ref{ass0}
where instead of demanding that for each $i \le m$ the map
$R^i_{\Sigma}:S^1 \times  [0,1] \to \Sigma$ is an injection
we only demand that $R^i_{\Sigma}(t_1,u_1) \neq R^i_{\Sigma}(t_2,u_2)$ for all $t_1, t_2 \in S^1$
and all $u_1, u_2 \in [0,1]$ fulfilling $u_1 \neq u_2$
and that for fixed $u \in [0,1]$ the Image of $S^1 \ni t \mapsto R^i_{\Sigma}(t,u) \in \Sigma$
lies on an embedded circle in $\Sigma$. \par

Observe that this includes a certain class of torus (ribbon) knots.
I expect that the obvious generalization of Conjecture  \ref{conj2} below will also hold
for the aforementioned weakened  version of Assumption \ref{ass0} (if combined with
a suitably modified version of Assumption \ref{ass1} below).
Moreover, I expect that Conjecture \ref{conj3} below can be generalized to this more general situation
and that by doing so one can obtain a  continuum analogue of Theorem 5.7 in \cite{Ha9}.
\end{remark}

\subsection{Definition of $\Det_{rig}(B)$}
\label{subsec2.5}

We will now explain how,
  using a suitable ``heat kernel regularization'', one can make rigorous sense
of the  expression
\begin{equation} \label{eq_heuristic_expr}
\Det(B) := \Det_{FP}(B)  \Check{Z}(B)
\end{equation}
and how one can evaluate the rigorous version $\Det_{rig}(B)$ of $\Det(B)$
 explicitly\footnote{in the special case where $B$ is constant
there is an additional way of doing so, cf. part ii) of Remark \ref{rm_sec2.5_2} below}.
Here $B \in \cB$ is fixed and  $\Det_{FB}(B)$ and $\Check{Z}(B)$ are given as in
 Eq. \eqref{eq_DetFP} and Eq. \eqref{eq_def_Z_B} above.

\begin{remark} \label{rm_sec2.5_1}
   The approach which we use here is a simplified
version of the approach in Sec. 6 in \cite{BlTh1}.
The main difference is that we use the exponentials  $e^{- \eps \triangle_i}$
of the original (=``plain'') Hodge Laplacians $\triangle_i$ while in Sec. 6 in \cite{BlTh1}
 ``covariant Hodge Laplacians'' are used.
The use of the covariant Hodge Laplacians produces an additional term
containing the  dual Coxeter number $\cg$ of $\cG$. The overall effect
in the simple situation in \cite{BlTh1} where only ``vertical links''
 (see the paragraph after Remark \ref{rm_sec2.5_2} below)
are used is a  ``shift'' $k \to k+\cg$, in agreement with the shift predicted in Witten's original paper \cite{Wi}.
In the case of general links it is doubtful that the use of
 covariant Hodge Laplacian can produce a shift $k \to k+\cg$ in all places where this would be necessary.
On the other hand, the fact that by working with the ``plain'' Hodge Laplacians
we do not get a shift $k \to k+\cg$  should not be a cause for concern.
It seems to be generally accepted  nowadays that
 the occurrence and magnitude of the  shift in $k$
will depend on the regularization procedure and renormalization prescription
 which is applied (cf. Remark 3.2 in \cite{Ha7a}).
  \end{remark}

Informally, we have
\begin{equation} \label{eq_sec2.5_1}
\Check{Z}(B) \sim |\det\bigl(\tfrac{\partial}{\partial t} + \ad(B)\bigr)\bigr|^{ -1/2} =
 \det\bigl(1_{\ck}-\exp(\ad(B))_{|\ck}\bigr)^{-1/2}
 \end{equation}
 where $\tfrac{\partial}{\partial t} + \ad(B)$ is as in Eq. \eqref{eq_SCS_expl} above
 and where   $1_{\ck}-\exp(\ad(B))_{|\ck}$  is the linear operator on $\cA_{\Sigma,\ck}=
 \Omega^1(\Sigma,\ck)$ given by
$$\forall  \alpha \in  \cA_{\Sigma,\ck}:  \forall  \sigma \in \Sigma:  \forall X_{\sigma} \in T_{\sigma} \Sigma:
\quad \quad
(1_{\ck}-\exp(\ad(B))_{|\ck} \cdot \alpha)(X_{\sigma}) = (\id_{\ck}-\exp(\ad(B(\sigma))_{|\ck}) \cdot \alpha(X_{\sigma})  $$
with $\id_{\ck}$  and $\exp(\ad(B(\sigma)))$  as in Sec. \ref{subsec2.2} above.

\smallskip

Now observe that for $b \in \ct$ we have (cf. Eq. \ref{eq_appA1} in Appendix \ref{appA} below)
\begin{equation} \label{eq_sec2.5_3}
\det(\id_{\ck}-\exp(\ad(b))_{|\ck}) =  \prod_{\alpha \in \cR+}  4 \sin^2(\pi  \alpha(b))
\end{equation}
where $\cR_+$ is  the set of positive real roots of $(\cG,\ct)$.\par

In view of Eqs. \eqref{eq_DetFP}, \eqref{eq_sec2.5_1}, and \eqref{eq_sec2.5_3}
 we now  rewrite the informal determinant $\Det(B)$ in Eq. \eqref{eq_heuristic_expr} as
$$ \Det(B) =  \prod_{\alpha \in \cR_+} \det(O^{(0)}_{\alpha}(B))^2 \det(O^{(1)}_{\alpha}(B))^{-1}$$
where  for each fixed $\alpha \in \cR_+$  the  operators
$O^{(i)}_{\alpha}(B): \Omega^i(\Sigma,\bR) \to \Omega^i(\Sigma,\bR)$, $i =0,1$,
 are the multiplication operators obtained by multiplication with the function
  $\Sigma \ni \sigma \mapsto 2 \sin(\pi  \alpha(B(\sigma))) \in \bR$.\par

Let us now equip the two spaces $\Omega^i(\Sigma,\bR)$, $i =0,1$, with
the scalar product which is induced by the Riemannian metric $\mathbf g$ on $\Sigma$  fixed in Sec. \ref{subsec2.2} above. By $\overline{\Omega^i(\Sigma,\bR)}$ we will denote the
completion of the pre-Hilbert space $\Omega^i(\Sigma,\bR)$, $i =0,1$. \par

Let us now define
\begin{subequations} \label{eq_Det_rig_defs}
\begin{equation} \label{eq_Det_rig_defsa} \Det_{rig}(B):=  \prod_{\alpha \in \cR_+} \Det_{rig,\alpha}(B)
\end{equation}
with
\begin{equation}\label{eq_Det_rig_defsb}  \Det_{rig,\alpha}(B):= \lim_{\eps \to 0} \bigl[ \det\nolimits_{\eps}(O^{(0)}_{\alpha}(B))^2 \det\nolimits_{\eps}(O^{(1)}_{\alpha}(B))^{-1} \bigr]
\end{equation}
where for $i=0,1$ we have set\footnote{this ansatz is, of course, motivated by the
rigorous formula $\det(A) = \exp(\Tr(\ln(A))$ which holds for every strictly positive
(self-adjoint) operator $A$ on a finite-dimensional Hilbert-space}
\begin{equation} \label{eq_Det_rig_defsc}
\det\nolimits_{\eps}(O^{(i)}_{\alpha}(B)) :=
 \exp\bigl( \Tr\bigl( e^{- \eps \triangle_i} \log(O^{(i)}_{\alpha}(B))\bigr) \bigr)
\end{equation}
\end{subequations}
Here $\triangle_i$ is the Hodge Laplacian on $\overline{\Omega^i(\Sigma,\bR)}$
  w.r.t. the
Riemannian metric  $\mathbf g$ on $\Sigma$ and
$\log: \bR \backslash \{0\} \to \bC$ is the restriction to  $\bR \backslash \{0\}$
 of the principal branch of the complex logarithm.
  We remark that in the special case  $B \in \cB_{reg}$, which we will assume in the following,
each of the bounded operators
$O^{(i)}_{\alpha}(B)$, $i = 0,1$, is a  symmetric operator
whose spectrum is bounded away from zero (cf. Eq. \eqref{eq_appA2} in Appendix \ref{appA})
 so    $\log(O^{(i)}_{\alpha}(B))$  is a well-defined bounded operator.
Moreover, since $e^{- \eps \triangle_i} $ is trace-class
 the product $e^{- \eps \triangle_i} \log(O^{(i)}_{\alpha}(B))$ is also trace-class and
the expression $\Tr( e^{- \eps \triangle_i} \log(O^{(i)}_{\alpha}(B)))$ is therefore well-defined.
Explicitly, we have
\begin{equation} \label{eq_heatkernel1}
\Tr( e^{- \eps \triangle_i} \log(O^{(i)}_{\alpha}(B))) = \int_{\Sigma} \Tr(K^{(i)}_{\eps}(\sigma,\sigma)) \log(2 \sin(\pi  \alpha(B(\sigma)))) d\mu_{\mathbf g}(\sigma)
\end{equation}
where  $K^{(0)}_{\eps}:\Sigma \times \Sigma \to \bR \cong \End(\bR)$
 is the integral kernel of $e^{- \eps \triangle_0}$
and $K^{(1)}_{\eps}:\Sigma \times \Sigma \to \bigcup_{\sigma_1,\sigma_2 \in \Sigma} \Hom(T_{\sigma_1} \Sigma, T_{\sigma_2} \Sigma)$ is the integral kernel of $e^{- \eps \triangle_1}$.
  According to a famous result in \cite{McSin}
  the  negative powers of $\eps$ that appear in the asymptotic
   expansion of $K^{(i)}_{\eps}$, $i =0,1$ as $\eps \to 0$ cancel each other  (= the ``fantastic cancelations'')
   and we obtain
\begin{equation} \label{eq_heatkernel_expr}
 \bigl[2 \Tr( K^{(0)}_{\eps}(\sigma,\sigma)) -   \Tr(K^{(1)}_{\eps}(\sigma,\sigma))\bigr]
 \to  \tfrac{1}{4 \pi}   R_{\mathbf g} (\sigma) \quad \text{ uniformly in $\sigma$ as $\eps \to 0$  }
 \end{equation}
 where $R_{\mathbf g}$ is the scalar curvature ( = twice the Gaussian curvature) of $(\Sigma,{\mathbf g})$. \par
  From Eqs. \eqref{eq_Det_rig_defsc}, \eqref{eq_heatkernel1}, and \eqref{eq_heatkernel_expr} it follows that
  the $\eps \to 0$ limit in Eq. \eqref{eq_Det_rig_defsb}  really exists and that we have
 \begin{equation} \label{eq_explicit_formula_Detreg} \Det_{rig,\alpha}(B) =
 \exp\biggl( \int_{\Sigma} \log(2 \sin(\pi  \alpha(B(\sigma))))  \tfrac{1}{4 \pi}   R_{\mathbf g} (\sigma) d\mu_{\mathbf g}(\sigma)\biggr)
\end{equation}

  In the special case where  $B \equiv b$  (with $b \in \ct_{reg}$)
 we can apply the classical  Gauss-Bonnet Theorem
 \begin{equation} \label{eq_Gauss_bonnet} 4 \pi \chi(\Sigma) =  \int_{\Sigma}      R_{\mathbf g}  d\mu_{\mathbf g}
 \end{equation}
  where $\chi(\Sigma)$ is the Euler characteristic of $\Sigma$
   and obtain
 \begin{equation} \Det_{rig,\alpha}(B) = (2 \sin(\pi  \alpha(b)))^{\chi(\Sigma)}
 \end{equation}
 and therefore
\begin{equation}\label{eq_sec2.5_B=b} \Det_{rig}(B) =  \bigl(\det\bigl(\id_{\ck}-\exp(\ad(b))_{| \ck}\bigr))^{\chi(\Sigma)/2}
 \end{equation}

So in particular, the value of $\Det_{rig}(B)$
is independent of the auxiliary Riemannian metric  ${\mathbf g}$ in this special case.

\begin{remark} \label{rm_sec2.5_2}
i) In the special case where $B$ is constant  the calculation of $\Det_{rig}(B)$
  just described can be simplified considerably.
  In particular, in this case  Eq. \eqref{eq_heatkernel_expr} is not necessary   for evaluating
  $\Det_{rig}(B)$ and proving Eq. \eqref{eq_sec2.5_B=b}.
  Indeed, for constant $B$ we only have to show that
  $\lim_{\eps \to 0}  \bigl(2 \Tr\bigl( e^{- \eps \triangle_0}) -  \Tr\bigl( e^{- \eps \triangle_1}) \bigr) = \chi(\Sigma)$. But this follows because
    $\lim_{\eps \to 0} \bigl( 2 \Tr\bigl( e^{- \eps \triangle_0}) -  \Tr\bigl( e^{- \eps \triangle_1}) \bigr)
    \overset{(*)}{=} 2 \dim(\ker(\triangle_0)) - \dim(\ker(\triangle_1))
   \overset{(**)}{=} 2 \dim(H^0(\Sigma,\bR)) -   \dim(H^1(\Sigma,\bR))
    =   \chi(\Sigma)$.
    Here step $(*)$ follows from another famous argument in \cite{McSin}
    and  step $(**)$ follows because according to the Hodge theorem we have $\Ker(\triangle_i) \cong H^i(\Sigma,\bR)$.

\smallskip

ii) We also mention that in the special case where $B$ is constant
there is an alternative way of defining and computing $\Det(B)$ using the Ray-Singer Torsion
(which makes use of a suitable $\zeta$-function regularization), cf.  Sec. 3 in \cite{BlTh1}.
\end{remark}

The special case $B \equiv b$  mentioned above was the only case which was relevant in \cite{BlTh1}
where only links consisting of ``vertical'' loops  were studied (at a heuristic level).
Here a ``vertical'' loop in $M= \Sigma \times S^1$ is a loop $l:S^1 \to \Sigma \times S^1$
which is ``parallel'' to $S^1$ (or in other words: $\arc(l_{\Sigma})$ is just a point in $\Sigma$).  \par

By contrast we will work with more general (ribbon) links
which means that  step functions $B$ of the type
\begin{equation} \label{eq_step_function} B= \sum_{i=1}^r b_i 1_{Y_i}, \quad \quad r \in \bN
\end{equation}
  will appear later during the explicit evaluation of $\WLO_{rig}(L)$  defined in Sec. \ref{subsec3.4} below,
  namely, after the $\eps \to 0$ and $s \to 0$-limits on the RHS of Eq. \eqref{eq_limit1}
  below have been carried out.
The regions $(Y_i)_{i \le r}$ here are the $r$ connected components of
\begin{equation} \Sigma \backslash \arc(L_{\Sigma}) = \Sigma \backslash \bigcup_{i=1}^m \arc(l^i_{\Sigma})
\end{equation}
In view of Eq. \eqref{eq_sec2.5_B=b} above one would expect that for $B:\Sigma \to \ct$ of the
form \eqref{eq_step_function} one has\footnote{observe that in contrast to $\chi(\Sigma)$, which is always
an even number,  $\chi(Y_i)$ can be odd, so in general we do {\it not} have
$\bigl(\det\nolimits^{1/2}\bigl(\id_{\ck}-\exp(\ad(b_i))_{| \ck}\bigr))^{\chi(Y_i)}
= \bigl(\det\bigl(\id_{\ck}-\exp(\ad(b_i))_{| \ck}\bigr))^{\chi(Y_i)/2}$}
\begin{equation} \label{eq_Detrig_step} \Det_{rig}(B) =  \prod_{i=1}^r \bigl(\det\nolimits^{1/2}\bigl(\id_{\ck}-\exp(\ad(b_i))_{| \ck}\bigr))^{\chi(Y_i)}
\end{equation}
where  $\det^{1/2}(\id_{\ck}-\exp(\ad(\cdot))_{|\ck}):\ct \to \bR$ is given by
\begin{equation} \label{eq_sec2.5_3b}
\det\nolimits^{1/2}(\id_{\ck}-\exp(\ad(b))_{|\ck}) =  \prod_{\alpha \in \cR+}  2 \sin(\pi  \alpha(b))
\end{equation}
And in fact, Eq. \eqref{eq_Detrig_step} is exactly the formula which is necessary for Conjecture \ref{conj3} below to be true.
The obvious question now is whether Eq. \eqref{eq_Detrig_step} follows from Eq. \eqref{eq_Det_rig_defsa} and  Eq. \eqref{eq_explicit_formula_Detreg} above\footnote{observe that the RHS of Eq. \eqref{eq_explicit_formula_Detreg} makes sense not only if $B: \Sigma \to \ct$ is smooth  but also if $B$ is  measurable, bounded, and bounded away from  $\ct_{sing} := \ct \backslash \ct_{reg}$,
 cf. Eq. \eqref{eq_appA2} in Appendix \ref{appA}.}.

 In order to answer this question recall
  the following, more general version\footnote{\label{ft_Gauss_Bonnet} In view of Remark \ref{rm_Sec2.5_Ende}  below we also remark that Eq. \eqref{eq_Gauss_bonnet_gen} can be generalized further to the situation where  the boundary $\partial Y$ is only a piecewise smooth
(rather than a smooth) submanifold of $\Sigma$. In the generalized formula there will be an extra term on the RHS involving a sum over the finite number of points $p$ of $\partial Y$ where
$\partial Y$ is not smooth (and containing the corresponding ``angle'' of  $\partial Y$ at $p$).}
 of  the classical Gauss-Bonnet Theorem mentioned above:
Let $Y \subset \Sigma$ be such that the boundary $\partial Y$ is (either empty or) a smooth 1-dimensional submanifold
of $\Sigma$. We equip $\partial Y$ with the Riemannian metric induced by $\mathbf g = \mathbf g_{\Sigma}$
and denote by $ds$ the corresponding ``line element''  on  $\partial Y$.
Then we have
\begin{equation} \label{eq_Gauss_bonnet_gen}
 4 \pi \chi(Y) =  \int_{Y}      R_{\mathbf g}  d\mu_{\mathbf g} + 2 \int_{\partial Y} k_{\mathbf g} ds
\end{equation}
where $k_{\mathbf g}(p)$ for $p \in \partial Y$ is the geodesic
 curvature of $\partial Y$ in the point $p$.\par

Let us now go back to the question
whether Eq. \eqref{eq_Detrig_step} follows from Eq. \eqref{eq_explicit_formula_Detreg}.
The short answer is: not for an arbitrary choice of $\mathbf g$ but for a natural subclass
of the possible choices, cf. Assumption \ref{ass1} and Remark \ref{rm_sec2.5_3} below.
Observe that in Eq. \eqref{eq_explicit_formula_Detreg}  there is no term involving geodesic curvature.
It is conceivable  that such a term appears  during the explicit evaluation
of the RHS of Eq. \eqref{eq_rm_step3_0} in Step 3 below as a result of ``self linking''.
However, even  in this case  the ``self linking'' expressions which we obtain will
 depend on the  precise regularization procedure which is used.
We plan to study this issue in more detail in the near future (cf. Question \ref{quest1} in Sec. \ref{sec4} below). In the present paper we will bypass this issue by
restricting ourselves to the situation where  the auxiliary Riemannian metric
 $\mathbf g $ chosen above fulfills a suitable condition.
 One sufficient condition would be to assume that $\mathbf g $ is chosen such that
 the geodesic curvature of each of the sets $\arc(l^i_{\Sigma})$ vanishes.
 In this case  Eq. \eqref{eq_Detrig_step} follows indeed from Eq. \eqref{eq_explicit_formula_Detreg}.
 However, since at first look
  this condition may seem somewhat unnatural
 we  will use the following assumption (which according to Remark \ref{rm_sec2.5_3} below is definitely natural)
 in the following:

\begin{assumption} \label{ass1} From now on we will assume that the auxiliary Riemannian metric
 $\mathbf g $ on $\Sigma$
was chosen such that
 on each  $\Image(\pi_{\Sigma} \circ R_i)$, $i \le m$,
 it coincides with the Riemannian metric
``induced'' by $\pi_{\Sigma} \circ R_i: S^1 \times (0,1) \to \Sigma$.
Here we have equipped $S^1 \times (0,1)$  with the product
of the  standard (normalized) Riemannian metrics on $S^1$ and on $(0,1)$.
\end{assumption}

We expect that  Assumption \ref{ass1} will lead to the correct values
for the rigorous implementation $\WLO_{rig}(L)$ of $\WLO(L)$, which we will give below,
 cf. Conjectures \ref{conj2} and \ref{conj3} in Sec. \ref{subsec3.4} below.
Moreover, the use of Assumption \ref{ass1}  eliminates
the regularization dependence of the ``self linking terms'' we referred to  above.

\begin{remark} \label{rm_sec2.5_3}
Recall that  the original heuristic path integral expression in Eq. \eqref{eq_WLO_orig} above is
 topologically invariant. In particular, it does not involve a  Riemannian metric.
 However,  for technical reasons, we  later introduced an auxiliary Riemannian metric  $\mathbf g$
 breaking topological invariance. \par

 Clearly, in a situation where one introduces an auxiliary object $\cO$ in order to make
 sense of a heuristic expression one of the following two principles (or a combination of them)
  ought to be fulfilled
 in order to be able to claim to have a natural treatment:
 \begin{enumerate}
 \item The auxiliary object   $\cO$  can be chosen arbitrarily
  and the final result does not depend on it.

 \item There is a distinguished/canonical choice of $\cO$ and that is the choice which we use.
\end{enumerate}

Our  Assumption \ref{ass1} is a combination of these two principles.
The restriction  of $\mathbf g$ on  $S := \bigcup_{i=1}^m
\Image(\pi_{\Sigma} \circ R_i)$ is given canonically.
 On the other hand, the restriction  of $\mathbf g$ on  $S^c:= \Sigma \backslash S$
 can be essentially\footnote{as long as $\mathbf g_{| S}$ and $\mathbf g_{|S^c}$ ``fit together'' smoothly,
  i.e. induce a smooth Riemannian metric on all of $\Sigma$}
  chosen arbitrarily.
\end{remark}

\begin{remark} \label{rm_Sec2.5_Ende} \rm  If one wants to study the situation of general (ribbon) links $L$,
i.e. links which need not fulfill Assumption \ref{ass0} above,
then  Assumption \ref{ass1} must be modified.
This is because when $U := \Image(\pi_{\Sigma} \circ R_i) \cap \Image(\pi_{\Sigma} \circ R_j) \neq \emptyset$
for $i \neq j$ then in general each of the two maps $ \pi_{\Sigma} \circ R_i$ and $\pi_{\Sigma} \circ R_j$
will induce a different Riemannian metric on $U$.
One way to deal with this complication is to
use, instead of Assumption \ref{ass1},
the aforementioned weaker condition\footnote{this condition
 is arguably quite natural as well  since it arises from   Assumption \ref{ass1}  (which is natural according to
 Remark \ref{rm_sec2.5_3} above)   by applying a suitable limit procedure} that  $\mathbf g $ is chosen such that
 the geodesic curvature of each of the sets $\arc(l^i_{\Sigma})$ vanishes.
Moreover, the generalization of Eq. \eqref{eq_Gauss_bonnet_gen} mentioned in Footnote \ref{ft_Gauss_Bonnet} above  will then be relevant.

\end{remark}

\subsection{The final heuristic formula (in the special case $\Sigma = S^2$)}
\label{subsec2.6}

Observe that
$$\cA^{\orth}_c \cong \cA_{\Sigma,\ct}$$
For simplicity we will assume in the following that $\Sigma \cong S^2$ and therefore
 $H^1(\Sigma)=\{0\}$.
In this case the Hodge decomposition of $\cA_{\Sigma,\ct}$ (w.r.t. the metric ${\mathbf g}$
fixed above) is given by
 \begin{equation} \label{eq_Hodge} \cA_{\Sigma,\ct} = \cA_{ex}  \oplus \cA^*_{ex}
 \end{equation}
 where
 \begin{align} \cA_{ex} & := \{d f \mid f \in C^{\infty}(\Sigma,\ct)  \}\\
 \cA^*_{ex} & := \{ \star d f \mid f \in C^{\infty}(\Sigma,\ct)\}
 \end{align}
 $\star$ being the relevant Hodge star operator.
  According to Eq. \eqref{eq_Hodge}
 we can  replace the
$\int \cdots DA_c^{\orth}$ integration in Eq. \eqref{eq2.48_ribbon}  by the integration
$ \int \int \cdots DA_{ex}  DA^*_{ex}$
where $DA_{ex}$, $DA^*_{ex}$ denote the heuristic ``Lebesgue measures''
on $\cA_{ex}$ and $\cA^*_{ex}$.\par

Taking this into account and replacing the heuristic expression $\Det(B)$
in Eq. \eqref{eq_heuristic_expr} by $\Det_{rig}(B)$ as given by Eqs \eqref{eq_Det_rig_defs}
we see that we can rewrite Eq. \eqref{eq2.48_ribbon} as
 \begin{multline} \label{eq_3.11}
\WLO(L)  \sim \lim_{s \to 0} \sum_{y \in I}
 \int_{\cA^*_{ex} \times C^{\infty}(\Sigma,\ct)} \biggl\{ \int_{\cA_{ex}}
   I^{(s)}(L)(A_{ex} + A^*_{ex} ,B) \bigr\}  DA_{ex}  \biggr\} \\
\times   \exp( -  2\pi i k   \langle y,  B(\sigma_0) \rangle  )
 1_{\cB_{reg}}(B)  \Det_{rig}(B) \\
  \times \exp(   2 \pi  i k  \ll\star  A^*_{ex},  dB \gg_{\cA^{\orth}})  (DA^*_{ex} \otimes    DB)
 \end{multline}
   where we have set
\begin{equation} \label{eq_ILs}
I^{(s)}(L)(A^{\orth}_c,B):=  \int_{\Check{\cA}^{\orth}}
  \prod_i \Tr_{\rho_i}(\Hol_{R^{(s)}_i}(\Check{A}^{\orth}, A^{\orth}_c,  B)
  \bigr)\bigr)    d\mu^{\orth}_B(\Check{A}^{\orth})
 \end{equation}
 for $A^{\orth}_c \in \cA^{\orth}_c \cong \cA_{\Sigma,\ct}$ and $B \in \cB$
and where we have used that
\begin{equation} \label{eq_L2-Gl}
  \ll\star  A^{\orth}_c,  d\Check{B} \gg_{\cA^{\orth}} =  \ll\star  A^{\orth}_c,  d\Check{B} \gg_{\cA_{\Sigma}} =
  - \ll \star dA^{\orth}_c ,\Check{B}\gg_{L^2_{\ct}(\Sigma,d\mu_{\mathbf g})}
  \end{equation}
 (which implies that $ \ll\star  A_{ex},  d\Check{B} \gg_{\cA^{\orth}} = 0$ if $A_{ex} \in \cA_{ex}$).

 \smallskip

 Using heuristic methods one can show\footnote{this is easy if $L$ fulfills Assumption \ref{ass0},
 cf. also Remark \ref{rm_Step3_1} below for a rigorous argument.
 For general (ribbon) links $L$ this point is not yet clear, cf. Sec. \ref{sec4} below}
\begin{equation} \label{eq_crucial} I^{(s)}(L)(A_{ex} + A^*_{ex} ,B) = I^{(s)}(L)( A^*_{ex} ,B)
\end{equation}
so, informally,
\begin{equation}\label{eq_3.12}
\int_{\cA_{ex}}   I^{(s)}(L)(A_{ex} + A^*_{ex} ,B)  \ DA_{ex} \sim  I^{(s)}(L)(A^*_{ex} ,B)
\end{equation}

Moreover, if we introduce the decomposition
 $ \cB = \Check{\cB} \oplus \cB_c$ where
\begin{align} \Check{\cB} & :=\{ B \in \cB \mid \int_{\Sigma} B \ d\mu_{\mathbf g}=0  \}\\
 \cB_c & :=\{ B \in \cB \mid B \text{ is constant} \} \cong \ct
\end{align}
we can also replace $\int \cdots DB$ by $ \int \int \cdots D\Check{B}  db$
where $D\Check{B}$ is the heuristic ``Lebesgue measure'' on $\Check{\cB}$
and $db$ is the (rigorous) normalized Lebesgue measure on $\cB_c \cong \ct$.
Taking this into account we obtain from Eqs. \eqref{eq_3.11} and \eqref{eq_3.12}
\begin{multline} \label{eq_WLO_FL_heuristic}
\WLO(L) \sim  \lim_{s \to 0}  \sum_{y \in I}  \biggl[ \int_{\cB_c} \ db \biggl[\int_{\cA^*_{ex} \times \Check{\cB}}
   I^{(s)}(L)(A^*_{ex} ,\Check{B}+b)
   \exp( -  2\pi i k \langle y,  \Check{B}(\sigma_0)+b \rangle   )\\
  1_{\cB_{reg}}(\Check{B}+b)
\Det_{rig}(\Check{B}+b) \\
 \times \exp(   - 2 \pi  i k  \ll\star  A^*_{ex},  d\Check{B} \gg_{\cA^{\orth}}) ( DA^*_{ex}\otimes D\Check{B})   \biggr] \biggr]
 \end{multline}
 Observe  that the operator
$\star d:\Check{\cB} \to \cA^*_{ex}$ is a linear isomorphism.
We can therefore make
 the change  of variable $\Check{B}_1: = (\star d)^{-1} A^*_{ex}$ and $\Check{B}_2:=\Check{B}$
 and rewrite Eq. \eqref{eq_WLO_FL_heuristic} as
\begin{equation} \label{eq_def_FL}
\WLO(L) \sim  \lim_{s \to 0}  \sum_{y \in I} \biggl[ \int_{\cB_c}  \biggl[ \int_{\Check{\cB} \times \Check{\cB}}
  J^{(s)}_{b,y}(L)(\Check{B}_1,\Check{B}_2) \  d\nu(\Check{B}_1,\Check{B}_2) \biggr] db \biggr]
 \end{equation}
 where we have set
 \begin{equation} \label{eq_def_JL}
 J^{(s)}_{b,y}(L)(\Check{B}_1,\Check{B}_2):= I^{(s)}(L)(\star d\Check{B}_1 ,\Check{B}_2+b)
   \exp( - 2\pi i k \langle y,  \Check{B}_2(\sigma_0) + b \rangle   )\\
  1_{\cB_{reg}}(\Check{B}_2+b)
\Det_{rig}(\Check{B}_2+b)
\end{equation}
and (cf. Eq. \eqref{eq_L2-Gl} above)
 \begin{equation} \label{eq_def_JL2} d\nu(\Check{B}_1,\Check{B}_2) := \tfrac{1}{Z}  \exp( - 2\pi i k
 \ll \star d \star d \Check{B}_1 ,\Check{B}_2\gg_{L^2_{\ct}(\Sigma,d\mu_{\mathbf g})})
   (D\Check{B}_1 \otimes D\Check{B}_2)
\end{equation}
with\footnote{Clearly, we could  drop the factor $1/Z$ in Eq. \eqref{eq_def_JL2}
since the symbol $\sim$ in Eq. \eqref{eq_def_FL}  denotes equality up to a multiplicative constant.
In Sec. \ref{subsec3.2}  we introduce a rigorous version of the integral functional
$\int \cdots d\nu$ and there it is convenient  that $d\nu$ is normalized.}
$$ Z:=  \int \exp( - 2\pi i k
 \ll \star d \star d \Check{B}_1 ,\Check{B}_2\gg_{L^2_{\ct}(\Sigma,d\mu_{\mathbf g})})
   (D\Check{B}_1 \otimes D\Check{B}_2)$$

\section{Rigorous realization of the RHS of Eq. \eqref{eq_def_FL}}
\label{sec3}

We will now explain how one can make rigorous sense of the  path integral expression appearing on the RHS
of Eq. \eqref{eq_def_FL} within the framework of White Noise Analysis.
In order to do so we will proceed in four steps:

\begin{description}
\item {\bf  Step 1:} We make rigorous sense of the integral functional $\int \cdots d\mu^{\orth}_B$
appearing in Eq. \eqref{eq_ILs}.

\item {\bf Step 2:}
We make rigorous sense of the integral functional $\int \cdots \ d\nu$ appearing in Eq. \eqref{eq_def_FL}.

\item {\bf  Step 3:}  We make rigorous sense of
the integral expression appearing in Eq. \eqref{eq_ILs} above, i.e. of
\begin{equation}
I^{(s)}(L)(A^{\orth}_c,B) =  \int_{\Check{\cA}^{\orth}}
  \prod_i \Tr_{\rho_i}(\Hol_{R^{(s)}_i}(\cdot, A^{\orth}_c,  B)
  \bigr)\bigr)    d\mu^{\orth}_B
 \end{equation}
for $A^{\orth}_c \in \cA^{\orth}_c$ and $B \in \cB$.

\item {\bf  Step 4:} We make rigorous sense of the total expression on the RHS of Eq. \eqref{eq_def_FL}.

\end{description}

\subsection{Step 1}
\label{subsec3.1}

We will now give a rigorous implementation
of  the integral functional $\int \cdots d{\mu}^{\orth}_B$ appearing in Eq.  \eqref{eq_ILs} above
as a  generalized distribution $\Phi^{\orth}_{B}$ on
a suitable extension
$\overline{\Check{\cA}^{\orth}}$ of $\Check{\cA}^{\orth}$.
\begin{enumerate}
\item First we will choose a suitable Gelfand triple $(\cN,\cH_{\cN},\cN')$
 and set $\overline{\Check{\cA}^{\orth}}:= \cN'$.\par

 Before we do this recall that
 \begin{align*}
 \cA^{\orth}  & \cong C^{\infty}(S^1,\cA_{\Sigma})\\
  \Check{\cA}^{\orth} & = \{ A^{\orth} \in \cA^{\orth} \mid \int A^{\orth}(t) dt \in \cA_{\Sigma,\ck} \}
 \end{align*}
 We set   $\cH_{\Sigma}:= L^2$-$\Gamma(\Hom(T\Sigma,\cG),d\mu_{\mathbf g})$,
i.e. $\cH_{\Sigma}$ is the Hilbert space
of $L^2$-sections (w.r.t. the measure $ d\mu_{\mathbf g}$)
  of the bundle $\Hom(T\Sigma,\cG) \cong T^{*}\Sigma \otimes \cG$
  equipped with  the fiber metric   induced by ${\mathbf g}$ and  $\langle \cdot,\cdot \rangle$.
  Moreover, we set\footnote{In other words: $\cH^{\orth}$ is the space of $\cH_{\Sigma}$-valued
(measurable) functions on $S^1$ which are square-integrable w.r.t. $dt$.}
  \begin{align*}
 \cH^{\orth}  & :=  L^2_{\cH_{\Sigma}}(S^1,dt)\\
  \Check{\cH}^{\orth} & := \{ H^{\orth} \in \cH^{\orth} \mid \int H^{\orth}(t) dt \in \cH_{\Sigma,\ck} \}
 \end{align*}
 where $\cH_{\Sigma,\ck}$  is the Hilbert space
 defined in a completely analogous way as $\cH_{\Sigma}$ but with $\ck$ playing the role of $\cG$.

 \medskip

 The Gelfand triple $(\cN,\cH_{\cN},\cN')$ we choose is  given by
\begin{align}
\cN & := \Check{\cA}^{\orth} \\
 \cH_{\cN} & := \Check{\cH}^{\orth}
\end{align}
where we have equipped $\Check{\cA}^{\orth}$ with a suitable\footnote{
more precisely, the family of semi-norms must be chosen such that
$\cN = \Check{\cA}^{\orth}$ is nuclear and  the inclusion map $\cN \to \cH_{\cN}$ is continuous}
  family of semi-norms.\par
Using ``second quantization''
and the Wiener-Ito-Segal isomorphism
$$ Fock_{sym}(\cH_{\cN}) \cong L^2_{\bC}(\cN',\gamma_{\cN'})$$
where $\gamma_{\cN'}$ is the canonical Gaussian measure  on
$\cN'$ (associated to  $(\cN,\cH_{\cN},\cN')$) we  obtain   a new Gelfand triple
$((\cN),L^2_{\bC}(\cN',\gamma_{\cN'}),(\cN)')$.

\item Next we   evaluate the Fourier transform
 $\cF {\mu}^{\orth}_B$  of the
informal measure ${\mu}^{\orth}_B$.  From Eq. \eqref{eq_SCS_expl} and Eq. \eqref{eq_def_mu_B}
we obtain immediately $$ \forall j \in \cN: \quad
\cF {\mu}^{\orth}_B (j)= \int \exp(i \ll \cdot,j \gg_{\cH_{\cN}})
 d{\mu}^{\orth}_B   =    \exp(- \tfrac{1}{2}  \ll j, C_B  j\gg_{\cH_{\cN}}) $$
where $C_B$  is given informally by
\begin{equation} \label{eq_def_CB} C_B =   \bigl(- 2 \pi k \star  ( \partial/\partial t + \ad(B))\bigr)^{-1} =
 \tfrac{1}{2 \pi k}  \star  ( \partial/\partial t + \ad(B))^{-1}
 \end{equation}

\smallskip

For each fixed $B \in \cB_{reg}$ we will now make sense of
 $( \partial/\partial t + \ad(B))^{-1}$ as  a densely defined
 linear operator on $\Check{\cH}^{\orth}$.
In order to do so we first introduce the
space $\Check{C}^{\infty}(S^1,\cG):= \{ f \in C^{\infty}(S^1,\cG) | \int f(t) dt \in \ck\}$.
It is not difficult to see that for
 $b \in \ct_{reg}$ the operator $\partial/\partial t + \ad(b): \Check{C}^{\infty}(S^1,\cG) \to
 \Check{C}^{\infty}(S^1,\cG)$
is invertible\footnote{Its inverse $(\partial/\partial t + \ad(b))^{-1}$
is given explicitly by (cf. Eq. (5.8) in \cite{Ha7a})
   \begin{equation} \label{eq_sec3.1_inv}  ((\partial/\partial t + \ad(b))^{-1} f)(t) =
T(b) \cdot \int_0^1 e^{s\ad(b)} f(t+i_{S^1}(s))  ds
 \end{equation}
 for all $f \in \Check{C}^{\infty}(S^1,\cG)$ and $t \in S^1$
 where   $i_{S^1}:[0,1] \ni s \mapsto e^{2 \pi i s} \in U(1) \cong S^1$
 and where $T(b) \in \End(\cG)$ is given by
 $T(b)(X) = (e^{\ad(b)}- \pi_{\ck})^{-1}(X)$ if $X \in \ck$
 and $T(b)(X) = X$ if $X \in \ct$.
  Here $\pi_{\ck}: \cG \to \ck$ is the orthogonal projection.
 We remark that in the special case where
$f$ takes only values in $\ct$  Eq. \eqref{eq_sec3.1_inv} reduces to $((\partial/\partial t + \ad(b))^{-1} f)(t) = ((\partial/\partial t)^{-1} f)(t) =   \int_0^1  f(t+i_{S^1}(s))  ds$}.

 Now let $(\partial/\partial t + \ad(B))^{-1}: \Check{\cA}   \to \Check{\cA}  \subset \Check{\cH}^{\orth}$
 be the linear operator  given by
\begin{equation} \label{eq_sec3.1_inv2}
\bigl((\partial/\partial t + \ad(B))^{-1} \cdot \Check{A}^{\orth})(t)\bigr)(X_{\sigma})
=  (\partial/\partial t + \ad(B(\sigma))^{-1} \cdot \bigl(\Check{A}^{\orth}(t)(X_{\sigma}) \bigr)
 \end{equation}
 for all $ \Check{A}^{\orth} \in  \Check{\cA}^{\orth}$,
  $t \in S^1$, $\sigma \in \Sigma$, and $X_{\sigma} \in T_{\sigma} \Sigma$.
 Observe that $(\partial/\partial t + \ad(B))^{-1} \cdot \Check{A}^{\orth}$ is indeed
 a well-defined element of $\Check{\cA}^{\orth}$ because, by the assumption on $B$ we have $B \in \cB_{reg}$,
 i.e.   $B(\sigma) \in \ct_{reg}$ for  all $\sigma \in \Sigma$. \par
It is easy to check that $(\partial/\partial t + \ad(B))^{-1}$ is  bounded and anti-symmetric.
Since $\star$ is  bounded and anti-symmetric  as well
we have now found a rigorous realization of the operator $C_B$  in Eq. \eqref{eq_def_CB}
as a  bounded  and symmetric operator on $\Check{\cH}^{\orth}$.

\smallskip

For technical reasons we need to define\footnote{observe that in Sec. \ref{subsec3.4} below we replace
the indicator function $1_{\cB_{reg}}$ by regularized versions
$1^{(n)}_{\cB_{reg}}$ and the condition $1^{(n)}_{\cB_{reg}}(B) \neq 0$ (where $n \in \bN$ is fixed)
does no longer guarantee  that $B \in \cB_{reg}$}  $C_B$ also for $B \notin \cB_{reg}$.
In view of the indicator function $1_{\cB_{reg}}$ appearing in Eq. \eqref{eq_limit1}
it seems\footnote{In fact, in spite of this indicator function $1_{\cB_{reg}}$
some of the functions $B$ which appear during the explicit evaluation
of the RHS of Eq. \eqref{eq_limit1}  will not be elements of $\cB_{reg}$ (for reasons explained
in the previous footnote).
It would be  more satisfactory to  modify our approach in a suitable way, for example by
using the regularization procedure described in Appendix \ref{appB} already now (and not only in Step 4),
 and working with $\Psi^{\orth}_{B^{(n)}}$ instead of $\Psi^{\orth}_{B}$
with $B^{(n)}$ given by Eq. \eqref{eq_def_Bn} below.
In the situation of ribbon links $L$ fulfilling Assumption \ref{ass0} above
 we can bypass this problem by simply defining $C_B$ by $C_B := 0$
 if $B \in \cB_{reg}$. With this choice Eq. \eqref{eq_rm_step3} in Proposition \ref{prop_3.2} will
hold for all $B \in \cB$ and we can expect Conjectures \ref{conj2} and \ref{conj3} to be true.}
that we are entitled to define $C_B$ in an arbitrary way if
 $B \notin \cB_{reg}$.  For simplicity we will take $C_B$ to be  trivial (i.e.
 $C_B = 0$) if $B \notin \cB_{reg}$.

\item For fixed $B \in \cB$ let $U_B:\cN \to \bC$ be the  well-defined continuous function  given by
\begin{equation}
U_B(j)=  \exp(- \tfrac{1}{2}  \ll j, C_B  j\gg_{\cH_{\cN}})
\end{equation}
for every $j \in \cN$.
It is straightforward to show that the function  $U_B:\cN \to \bC$
   is a ``$U$-functional'' in the sense
 of \cite{HKPS,KLPSW}. In view of the Kondratiev-Potthoff-Streit Characterization Theorem
 (cf. again \cite{HKPS,KLPSW}) the integral functional $\Phi^{\orth}_{B}:= \int \cdots d{\mu}^{\orth}_B$
can  be defined rigorously as
the unique element  $\Phi^{\orth}_{B}$ of $(\cN)'$
such that
\begin{equation} \Phi^{\orth}_{B}(\exp(i(\cdot,j)_{\cN}))  = U_B(j)
\end{equation}
holds for all $j \in \cN$.
Here $(\cdot,\cdot)_{\cN}:\cN' \times \cN \to \bR$ is the canonical pairing.
\end{enumerate}

\begin{convention} \label{conv1}
Let $\pi: \cA^{\orth} = \Check{\cA}^{\orth} \oplus \cA^{\orth}_c \to \Check{\cA}^{\orth}$
be the canonical projection. The map
 $\pi': (\Check{\cA}^{\orth})' \to (\cA^{\orth})'$ which is dual to $\pi$ is an injection.
Using $\pi'$ we will identify $(\Check{\cA}^{\orth})'$ with a subspace of $(\cA^{\orth})'$. \par

 Moreover, we will identify each element $A^{\orth}$ of $(\cA^{\orth})'$
 with the continuous map $f_{A^{\orth}}: \cA^{\orth}_{\bR} \to \cG$ given by
$f_{A^{\orth}}(\psi^{\orth}) = \sum_a T_a (A^{\orth}, T_a \psi^{\orth})$ for all $ \psi^{\orth}
\in \cA^{\orth}_{\bR}$
where $\cA^{\orth}_{\bR}:= C^{\infty}(S^1,\cA_{\Sigma,\bR})$ and
  where $(T_a)_a$ is any fixed $\langle \cdot, \cdot \rangle$-orthonormal basis of $\cG$.
\end{convention}

\subsection{Step 2}
\label{subsec3.2}

In order to make rigorous sense
of the heuristic integral functional $ \int_{\Check{\cB} \times \Check{\cB}} \cdots d\nu$
as a generalized distribution on a suitable extension
$ \overline{\Check{\cB} \times \Check{\cB}}$ of the space $\Check{\cB} \times \Check{\cB}$
we will proceed in a similar way as in Step 1 above.

\begin{enumerate}
\item First we choose a suitable Gelfand triple
$(\cE,\cH_{\cE},\cE')$
and set $$\overline{\Check{\cB} \times \Check{\cB}}:= \cE'.$$
More precisely, we choose
\begin{align}
\cE & :=  \Check{\cB} \times \Check{\cB}\\
\cH_{\cE} & := \Check{L}^2_{\ct}(\Sigma,d\mu_{\mathbf g}) \oplus \Check{L}^2_{\ct}(\Sigma,d\mu_{\mathbf g})
\end{align}
where we have equipped $\cE$
with a suitable family of semi-norms and where we have set $\Check{L}^2_{\ct}(\Sigma,d\mu_{\mathbf g}):=
 \{f \in L^2_{\ct}(\Sigma,d\mu_{\mathbf g}) \mid \int f d\mu_{\mathbf g} = 0\}$.
 Using second quantization and the Wiener-Ito-Segal isomorphism
$$ Fock_{sym}(\cH_{\cE}) \cong L^2_{\bC}(\cE',\gamma_{\cE'})$$
 where $\gamma_{\cE'}$ is the canonical Gaussian measure on $\cE'$ (associated to $(\cE,\cH_{\cE},\cE')$)
we obtain   a new Gelfand triple $((\cE),L^2_{\bC}(\cE',\gamma_{\cE'}),(\cE)')$.

\item Next we evaluate the Fourier transform $\cF \nu$   of the heuristic
``measure'' $\nu$ at an informal level.
Clearly,
 $\nu$ is of ``Gauss type'' with  the well-defined\footnote{observe that
 the kernel of $\triangle: \cB \to \cB$ equals $\cB_c$, so $\triangle_{| \Check{\cB}}$ is
 injective  } covariance operator
 $$C:= -\tfrac{1}{ 2\pi k} \biggl( \begin{matrix}  0 & (\triangle_{| \Check{\cB}})^{-1}\\  (\triangle_{| \Check{\cB}})^{-1} & 0
\end{matrix} \biggr), \quad \quad \text{where } \triangle:= \star d \star d$$

Taking this into account we obtain
$$\forall j \in \cE: \quad
\cF \nu(j) = \int \exp(i \ll \cdot,j \gg_{\cH_{\cE}}) d\nu =
 \exp(- \tfrac{1}{2}  \ll j, C  j\gg_{\cH_{\cE}})$$
We remark that  $C$ is a (densely defined)
bounded and symmetric linear operator on $\cH_{\cE}$.

\item Let $U:\cE \to \bC$ be given by
\begin{equation}
U(j)=  \exp(- \tfrac{1}{2}  \ll j, C  j\gg_{\cH_{\cE}})
\end{equation} for every $j \in \cE$.
Clearly,   $U:\cE \to \bC$    is a ``$U$-functional'' in the sense
 of \cite{HKPS,KLPSW} so using the Kondratiev-Potthoff-Streit Characterization Theorem
the integral functional
 $\Psi := \int_{\Check{\cB} \times \Check{\cB}} \cdots d\nu$
can  be defined rigorously as
the unique element  $\Psi$ of $(\cE)'$
such that
\begin{equation}\Psi(\exp(i(\cdot,j)_{\cE}))  = U(j)
\end{equation}
holds for all $j \in \cE$.
Here $(\cdot,\cdot)_{\cE}:\cE' \times \cE \to \bR$ is the canonical pairing.
\end{enumerate}

\begin{convention} \label{conv2}
Let $\pi: \cB = \Check{\cB} \oplus \cB_c \to \Check{\cB}$ be the canonical projection.
The map  $\pi': \Check{\cB}' \to \cB'$ which is dual to $\pi$
is an injection. Using $\pi'$ we will identify $\Check{\cB}'$ with a subspace of $\cB'$. \par

Moreover, we will identify each element $B$ of $\cB'$ with the continuous map
 $f_{B}: C^{\infty}(\Sigma,\bR) \to \ct$ given by
$f_{B}(\psi) = \sum_a T_a (B, T_a \psi)_{\cB}$ for all $\psi \in C^{\infty}(\Sigma,\bR)$
where $(\cdot,\cdot)$ is the canonical pairing $\cB \times \cB' \to \bR$
and  $(T_a)_a$ is a fixed orthonormal basis of $\ct$.

\end{convention}

\subsection{Step 3}
\label{subsec3.3}

Let us now make rigorous sense of
the heuristic integral
\begin{equation}
I^{(s)}(L)(A^{\orth}_c,B) = \int_{ \Check{\cA}^{\orth} }  \prod_i \Tr_{\rho_i}(\Hol_{R^{(s)}_i}(\Check{A}^{\orth},
 A^{\orth}_c,  B)
  \bigr)\bigr)    d\mu^{\orth}_B(\Check{A}^{\orth})
 \end{equation}
 for $A^{\orth}_c \in \cA^{\orth}_c$ and $B \in \cB$.
 We already have a rigorous version
$\Phi^{\orth}_{B}$ of the heuristic integral functional
 $\int \cdots d{\mu}^{\orth}_B$.
 However, clearly we can not just consider
 $$\Phi^{\orth}_{B}\bigl(
 \prod_i \Tr_{\rho_i}\bigl(\Hol_{R^{(s)}_i}(\cdot, A^{\orth}_c,  B)
   \bigr)\bigr)$$
 since the function
$\Hol_{R^{(s)}_i}(\cdot, A^{\orth}_c,  B)$
  was defined as a function on   $\Check{\cA}^{\orth}$
  and not as a function on all of
  $\overline{\Check{\cA}^{\orth}} = \cN' $.

 Let us now fix  $j \le m$ and $s > 0$ temporarily  (until Proposition \ref{prop_3.1})
 and set $R:= R^{(s)}_j$ .
 Using\footnote{of course, we also have $A^{\orth}_c((R_u)'(t)) = A^{\orth}_c((\pi_{\Sigma} \circ R_u)'(t))$
 but we will not need this}
  $\Check{A}^{\orth}((R_u)'(t)) = \Check{A}^{\orth}((\pi_{\Sigma} \circ R_u)'(t))$
 we obtain
 \begin{align} \label{eq_sec3.3_1}
  & \Hol_{R}(\Check{A}^{\orth}, A^{\orth}_c,  B) \nonumber \\
 & =  \lim_{n \to \infty} \prod_{j=1}^n \exp\bigl(\tfrac{1}{n} \int_{0}^1  \bigl[
 (\Check{A}^{\orth} +  A^{\orth}_c  +  B dt)(R'_u(t)) \bigr] du \bigr)_{| t=j/n} \nonumber \\
 & =  \lim_{n \to \infty} \prod_{j=1}^n \exp\bigl(\tfrac{1}{n} \int_{0}^1  \bigl[
 \Check{A}^{\orth}((\pi_{\Sigma} \circ R_u)'(t))  +  (A^{\orth}_c  +  B dt)(R'_u(t))  \bigr] du \bigr)_{| t=j/n}
   \end{align}
  Clearly, for   a general element $\Check{A}^{\orth}$ of $\overline{\Check{\cA}^{\orth}} = \cN'$
  the expression  $ \Check{A}^{\orth}((\pi_{\Sigma} \circ R_u)'(t)) $ appearing in the last expression
  does not make sense. \par

  In order to get round this complication we will now make use of ``point smearing'', i.e. replace
  points by suitable test functions.

  \smallskip

  In order to do so we  choose, for each $\sigma \in \Sigma$ a  Dirac family\footnote{i.e.  for each fixed $\sigma \in \Sigma$  we have the following:
  $\delta^{\eps}_{\sigma}$, $\eps >0$, is a non-negative and smooth function  $\Sigma \to \bR$. Moreover,
 $\int \delta^{\eps}_{\sigma} d\mu_{\mathbf g}=1$,
and we have $\delta^{\eps}_{\sigma} \to  \delta_{\sigma}$ weakly as $\eps \to 0$ where
$\delta_{\sigma}$ is the Dirac distribution in the point $\sigma$} $(\delta^{\eps}_{\sigma})_{\eps>0}$
around $\sigma$ w.r.t. $d\mu_{\mathbf g}$.
Moreover, for each $t \in S^1$ we choose a Dirac family $ (\delta^{\eps}_{t})_{\eps>0}$
around $t$ w.r.t. the measure $dt$ on $S^1$.\par

For every  $p = (\sigma,t) \in  \Sigma \times S^1$ and $\eps>0$ we define
$\delta^{\eps}_p \in C^{\infty}(\Sigma \times S^1,\bR)$ by
$$\delta^{\eps}_p(\sigma',t'):= \delta^{\eps}_{\sigma}(\sigma') \delta^{\eps}_{t}(t') \quad \text{
for all $\sigma' \in \Sigma$ and $t' \in S^1$.}$$

For technical reasons  we will  assume\footnote{in view of Question \ref{quest3} in Sec. \ref{sec4} below
we remark that if we want to treat the case of general ribbon links $L$ (i.e. ribbon links for which  Assumption \ref{ass0} need  not be fulfilled) we will probably
  have to make some additional technical assumptions on the family $(\delta^{\eps}_{\sigma})_{\sigma \in \Sigma}$}
   also  that for each fixed   $\eps > 0$
the family   $(\delta^{\eps}_{\sigma})_{\sigma \in \Sigma}$ was chosen
  such that    the function  $\Sigma \times \Sigma \ni (\sigma,\bar{\sigma}) \to \delta^{\eps}_{\sigma}(\bar{\sigma}) \in \bR$ is smooth
   and, moreover, to have the property  that
  for each $\eps$ and $\sigma \in \Sigma$ the support of  $\delta^{\eps}_{\sigma}$
  is contained in the $\eps$-ball w.r.t. $d_{\mathbf g}$ around $\sigma$.

\smallskip

Recall that $R= R^{(s)}_j$, $j \le m$, $s > 0$.
Let us now also introduce the notation $\bar{R}:= R_j$.
Let  $\eps_j(s)$ be the supremum of all $\eps>0$
such that for all $t \in S^1$,
and $u \in [0,1]$ we have\footnote{observe that  $\delta^{\eps}_{R_u(t)}$
depends on $s$ even though this is not reflected in the notation}
  $\supp( \delta^{\eps}_{R_u(t)}) \subset \bar{R}_{\Sigma}$.
Let $X_{\bar{R}_{\Sigma}}$ be the vector field on $\Image(\bar{R}_{\Sigma}) \subset \Sigma$,
which is induced by the collection of loops $S^1 \ni t \mapsto \bar{R}_{\Sigma}(t,u) \in \Sigma$, $ u \in [0,1]$. \par

After these preparations we can  now introduce  ``smeared'' analogues for the expression
 $ \Check{A}^{\orth}((\pi_{\Sigma} \circ R_u)'(t))$ appearing in Eq. \eqref{eq_sec3.3_1} above.
 More precisely, we now replace, for fixed $\eps \in (0,\eps_j(s))$ the expression
$ \Check{A}^{\orth}((\pi_{\Sigma} \circ R_u)'(t))$
 by the expression $\Check{A}^{\orth}(X_{\bar{R}_{\Sigma}} \delta^{\eps}_{R_u(t)})$.
 Here we made the identification  $VF(\Sigma) \cong \cA_{\Sigma,\bR}$
 using the Riemannian metric ${\mathbf g}$.
 On the other hand $ \cA_{\Sigma,\bR} \subset C^{\infty}(S^1, \cA_{\Sigma,\bR})
 \cong \cA^{\orth}_{\bR} $, so according to Convention \ref{conv1} above  the expression
  $\Check{A}^{\orth}(X_{\bar{R}_{\Sigma}} \delta^{\eps}_{R_u(t)})$
  is  well-defined for every $\Check{A}^{\orth} \in \overline{\Check{\cA}^{\orth}} = \cN'$.
 We can now set
   \begin{multline}
   \Hol^{(\eps)}_{R}(\Check{A}^{\orth}, A^{\orth}_c,  B) := \\
 \lim_{n \to \infty} \prod_{j=1}^n \exp\bigl(\tfrac{1}{n} \int_{0}^1  \bigl[
\Check{A}^{\orth}(X_{\bar{R}_{\Sigma}} \delta^{\eps}_{R_u(t)})  + (A^{\orth}_c  +  B dt)(R'_u(t))
 \bigr] du \bigr)_{| t=j/n}
 \in G \end{multline}

Using similar methods as in the proof of Proposition 6 in \cite{Ha2}
 it is not  difficult to prove the following result:

\begin{proposition} \label{prop_3.1}  For every $s \in (0,1)$ and every $\eps \in (0,\eps(s))$ where $\eps(s):= \min_{i \le m} \eps_i(s)$ we have
\begin{equation}
\prod_i \Tr_{\rho_i}\bigl(\Hol^{(\eps)}_{R^{(s)}_i}(\cdot, A^{\orth}_c,  B)   \bigr) \in (\cN)
\end{equation}
Consequently, the expression
\begin{equation} \label{eq_rm_step3_0}
I^{(s,\eps)}_{rig}(L)(A^{\orth}_c,B) := \Phi^{\orth}_{B}\bigl(\prod_i \Tr_{\rho_i}\bigl(\Hol^{(\eps)}_{R^{(s)}_i}(\cdot, A^{\orth}_c,  B)   \bigr)\bigr)
\end{equation}
is well-defined.
\end{proposition}

\begin{proposition} \label{prop_3.2}
 For every  $L$  fulfilling Assumption \ref{ass0} above
we have\footnote{recall that we also assume that Assumption \ref{ass1}
is fulfilled}
\begin{equation} \label{eq_rm_step3}
I^{(s,\eps)}_{rig}(L)(A^{\orth}_c,B) 
= \prod_i \Tr_{\rho_i}\biggl(\exp\biggl(\int_{0}^1 \biggl(  \int_{(R^{(s)}_i)_u}  ( A^{\orth}_c + Bdt )
\biggr) du \biggr)\biggr)
\end{equation}
where for a loop $l:S^1 \to M$ and a 1-form $\alpha$ on $M$ we use the  notation
$\int_l \alpha = \int_{S^1} l^*(\alpha)$.
Observe  that the RHS of Eq. \eqref{eq_rm_step3}
 actually does not depend on $\eps$.
\end{proposition}

\begin{remark} \label{rm_Step3_1}
We remark that Eq. \eqref{eq_rm_step3} implies  that
$I^{(s,\eps)}_{rig}(L)(A_{ex}+A^{*}_{ex},B) = I^{(s,\eps)}_{rig}(L)(A^{*}_{ex},B)$
(cf. the notation in Sec. \ref{subsec2.6} above)
which can be seen as a rigorous  justification
 of Eq. \eqref{eq_crucial} in Sec. \ref{subsec2.6} above.
\end{remark}

\begin{remark} \label{rm_Step3_Ende} \rm
\begin{enumerate}
\item Here is an alternative way for defining a
 ``smeared'' analogue of the expression  $ \Check{A}^{\orth}((\pi_{\Sigma} \circ R_u)'(t))$.
Observe that since $\Sigma$ is compact there is a $\eps_0 > 0$
such that for all $\sigma_0, \sigma_1 \in \Sigma$ with $d_{\mathbf g}(\sigma_0,\sigma_1) < \eps_0$ there is a unique
(geodesic) segment starting in $\sigma_0$ and ending in $\sigma_1$.
Using parallel transport along this geodesic segment w.r.t. the Levi-Civita connection of $(\Sigma,\mathbf g)$
we can transport every tangent vector $v \in T_{\sigma_0} \Sigma$
to a tangent vector in  $T_{\sigma_1} \Sigma$.
Thus every $v \in T_{\sigma_0} \Sigma$ induces in a natural way a vector field $X_v$
on the open ball $B_{\eps_0}(\sigma_0) \subset \Sigma$. \par
For every  $\eps < \eps_0$ we replace $ \Check{A}^{\orth}((\pi_{\Sigma} \circ R_u)'(t))$
 by the expression $\Check{A}^{\orth}(X_{(\pi_{\Sigma} \circ R_u)'(t)} \delta^{\eps}_{R_u(t)})$ .

\item The alternative method also works when $L$ does not fulfill Assumption \ref{ass0}
while the original method must be modified (in a relatively straightforward way) if
$L$ does not fulfill Assumption \ref{ass0}.

\item When both Assumption \ref{ass0} and Assumption \ref{ass1} are fulfilled
then both methods described here are equivalent.
Indeed, for each fixed $t$ and $u$ the vector field
$X_{(\pi_{\Sigma} \circ R_u)'(t)}$
coincides with the vector field $X_{\bar{R}_{\Sigma}}$ on the subset $S \subset \Sigma$
where both vector fields are defined.
For sufficiently small $\eps > 0$ we therefore have
$ X_{(\pi_{\Sigma} \circ R_u)'(t)} \delta^{\eps}_{R_u(t)} =   X_{\bar{R}_{\Sigma}} \delta^{\eps}_{R_u(t)}$.

 \item If Assumption \ref{ass1} is not fulfilled then the two methods described here
will probably not be equivalent. In particular, it seems that non-trivial self-linking terms
appear when using the  alternative method while no such self-linking term
will arise when the original method is used.
 \end{enumerate}
\end{remark}

\subsection{Step 4}
\label{subsec3.4}

Finally, let us make rigorous sense of
the full heuristic expressions on the RHS
of Eq. \eqref{eq_def_FL} above. \par

For similar reasons as in Step 3 above we will use again ``point smearing''.
Recall that above we chose for each $\sigma \in \Sigma$  a  ``Dirac family'' $(\delta^{\eps}_{\sigma})_{\eps>0}$
such that  for every $\eps > 0$ the function
 $\Sigma \times \Sigma \ni (\sigma,\bar{\sigma}) \to \delta^{\eps}_{\sigma}(\bar{\sigma}) \in \bR$ is smooth.
This implies\footnote{In the special case   $B=f \cdot d\mu_{\mathbf g}$
where $f: \Sigma \to \ct$ is continuous the smoothness of  $B^{(\eps)}: \Sigma \to \ct$  follows easily from
the assumption that
$\Sigma \times \Sigma \ni (\sigma,\bar{\sigma}) \mapsto \delta^{\eps}_{\sigma}(\bar{\sigma}) \in \bR$
is smooth. Moreover, the smoothness of $B^{(\eps)}$
 follows also if $B$ is any derivative of a distribution of the form $f \cdot d\mu_{\mathbf g}$ with
 $f \in C^0(\Sigma,\ct)$.
This covers already the general situation since, by a
 well-known theorem, every distribution $D \in \cD'(\Sigma)$
    can be written as a linear combination of derivatives of distributions of the form
 $f \cdot d\mu_{\mathbf g}$ where $f$ is a continuous function $\Sigma \to \bR$ and this result can  immediately
be generalized to the case of $\ct$-valued functions and distributions}
  that for each fixed $\eps>0$ and each $B \in  \cB'$
the function
$B^{(\eps)}: \Sigma \to \ct$ given by $B^{(\eps)}(\sigma) =  B(\delta^{\eps}_{\sigma})$ for all $\sigma \in \Sigma$
 (cf. Convention \ref{conv2} above) is smooth.
Consequently, the function
$\Check{B}^{(\eps)}: \Sigma \to \ct$ given by
$\Check{B}^{(\eps)}= B^{(\eps)} - \int B^{(\eps)}  d\mu_{\mathbf g}$
is a well-defined element of $\Check{\cB}$.
(For  $\Check{B} \in \Check{\cB}' \subset \cB'$ we will simply write $\Check{B}^{(\eps)}$
instead of $\Check{\Check{B}}^{(\eps)}$.)

\smallskip

For fixed $y \in I$, $b \in \ct$,  $s \in (0,1)$, and $\eps \in (0,\eps(s))$ we could now introduce
the function $J^{(s,\eps)}_{b,y}(L) : \cE' \to \bR$ by
\begin{multline}J^{(s,\eps)}_{b,y}(L)(\Check{B}_1,\Check{B}_2):= I^{(s,\eps)}_{rig}(L)(\star d\Check{B}_1^{(\eps)} ,\Check{B}_2^{(\eps)} +b) \cdot
 \exp( - 2\pi i k \langle y,  \Check{B}_2^{(\eps)}(\sigma_0) + b \rangle  )\\
 \times  \Det_{rig}(\Check{B}_2^{(\eps)}+b) 1_{\cB_{reg}}(\Check{B}_2^{(\eps)} + b)
\end{multline}
for all $(\Check{B}_1,\Check{B}_2) \in  (\Check{\cB})' \times (\Check{\cB})' \cong \cE'$.\par

However, the last two factors $\Det_{rig}(\Check{B}_2^{(\eps)}+b)$ and
 $1_{\cB_{reg}}(\Check{B}_2^{(\eps)} + b)$
are problematic since neither of these two factors (considered as  functions  $\cE' \to \bR$)
is an  element of $(\cE)$.

\smallskip

This is why, in addition to ``point smearing'', we will use an additional regularization and
introduce regularized versions
  $ 1^{(n)}_{\cB_{reg}}: \cB \to \bR$
  and  $\Det^{(n)}_{rig}: \cB \to \bR$,  $n \in \bN$,
  of $1_{\cB_{reg}}$ and $\Det_{rig}$.
 There are several ways to do this.
   In Appendix \ref{appB} we explain one possible regularization.
The following definitions, results, and conjectures
refer to the choice of  $ 1^{(n)}_{\cB_{reg}}: \cB \to \bR$
  and  $\Det^{(n)}_{rig}: \cB \to \bR$  of Appendix \ref{appB}.

  \medskip

Let $y \in I$, $b \in \ct$,   $s \in (0,1)$, $\eps \in (0,\eps(s))$,  and $n \in \bN$ be fixed.
We introduce
the function $J^{(s,\eps,n)}_{b,y}(L) : \cE' \to \bR$ by
\begin{multline}J^{(s,\eps,n)}_{b,y}(L)(\Check{B}_1,\Check{B}_2):= I^{(s,\eps)}_{rig}(L)(\star d\Check{B}_1^{(\eps)} ,\Check{B}_2^{(\eps)} +b) \cdot
 \exp( - 2\pi i k \langle y,  \Check{B}_2^{(\eps)}(\sigma_0) + b\rangle  )\\
 \times  \Det^{(n)}_{rig}(\Check{B}_2^{(\eps)}+b) 1^{(n)}_{\cB_{reg}}(\Check{B}_2^{(\eps)} + b)
\end{multline}
for all $(\Check{B}_1,\Check{B}_2) \in  (\Check{\cB})' \times (\Check{\cB})' \cong \cE'$.

\begin{proposition} \label{prop_conj1}  For all $b \in \ct$,  $y \in I$, $s \in (0,1)$, $\eps \in (0,\eps(s))$,
 and $n \in \bN$  we have
$$ J^{(s,\eps,n)}_{b,y}(L) \in (\cE)$$
Consequently, the expression $\Psi\bigl(
J^{(s,\eps,n)}_{b,y}(L)\bigr)$ is well-defined.
\end{proposition}

 After these preparations we can finally write down a rigorous version of the
heuristic expression on the RHS of Eq. \eqref{eq_def_FL} above:
\begin{equation} \label{eq_limit1}
 \WLO_{rig}(L) := \lim_{n \to \infty}   \lim_{s \to 0}  \lim_{\eps\to 0}
 \sum_{y \in I} \int_{\sim}     \Psi\bigl(
J^{(s,\eps,n)}_{b,y}(L)\bigr)   db
\end{equation}
where\footnote{recall that
$db$ is the normalized Lebesgue measure on $\cB_c \cong \ct$.
We expect  that the function $\cB_c \ni b \mapsto   \Psi\bigl(
J^{(s,\eps,n)}_{b,y}(L)\bigr) \in \bR$ appearing in Eq.  \eqref{eq_limit1} is periodic.
This is why instead of using the proper Lebesgue integral
$\int \cdots db$ we use the ``mean value'' $\int_{\sim} \cdots db$} $\int_{\sim} \cdots db$ is given by
$$\int_{\sim} f db  = \lim_{T \to \infty} \tfrac{1}{(2T)^d} \int_{[-T,T]^d } f db$$
for any measurable bounded function $f: \ct \to \bR$.
Here  $d = \dim(\ct)$ and  we have identified  $\ct$ with $\bR^d$ using any fixed orthonormal basis
$(e_i)_{i \le d}$ of $\ct$.

\begin{conjecture} \label{conj2}
 $\WLO_{rig}(L)$ is well-defined. In particular,  all limits involved exist.
\end{conjecture}

If Conjecture \ref{conj2} above is correct then
in view of the semi-rigorous computations in \cite{Ha4,HaHa} (and the rigorous computations in \cite{Ha7a,Ha7b})
 one naturally arrives at the following conjecture:

\begin{conjecture} \label{conj3} Assume that $k > \cg$ where $\cg \in \bN$ is the dual Coxeter number of $\cG$
(cf. Appendix \ref{appA}).
Then we have for every  $L$
fulfilling Assumption \ref{ass0} above
\begin{equation} \label{eq_theorem} \WLO_{rig}(L) \sim |L|
\end{equation}
where $\sim$ denotes equality up to a multiplicative constant $C = C(G,k)$
and where  $|\cdot|$ is the shadow invariant for $M=S^2 \times S^1$
associated to the pair $(\cG,k)$, cf. Appendix \ref{appA} for the definitions and concrete
formulas (cf., in particular, Eq. \ref{eqA.8}).
 \end{conjecture}

\begin{remark}
\begin{enumerate}
\item  As the notation $C = C(G,k)$ suggests the constant $C$ referred to above
is allowed to depend on $G$ and $k$ but will be independent of $L$.
It will also be independent of the particular choice of the  orthonormal basis
$(e_i)_i$ of $\ct$ and the Dirac families $\{ \delta^{\eps}_{\sigma} \mid \eps >0, \sigma \in \Sigma\}$ and
 $\{ \delta^{\eps}_{t} \mid \eps>0, t \in S^1\}$ above.
 Finally, it will be independent of the particular choice of the auxiliary Riemannian metric
 $\mathbf g$ (as long as $\mathbf g$ fulfills Assumption \ref{ass1} above).

\item Obviously we cannot expect Eq. \eqref{eq_theorem} to hold with ``$\sim$'' replaced
by ``$=$'' since in Sec. \ref{sec2} we have  omitted several multiplicative constants.
Moreover, Eq. \eqref{eq_def_FL} contains ``$\sim$'' as well.

\item In the standard literature the shadow invariant $| \cdot |$
associated to the pair $(\cG,k)$ is only defined when $k > \cg$.
It can easily be generalized in a natural way  so that it includes the situation $k \le \cg$
but it turns out that the so defined generalization of  $| \cdot |$ vanishes  for $k < \cg$
and is essentially trivial for $k = \cg$.
We expect that the same applies to  $\WLO_{rig}(L)$.
\end{enumerate}
\end{remark}

\begin{remark}
Recall from Remark \ref{rm_ass0} above that I expect that
both Conjecture  \ref{conj2} and Conjecture \ref{conj3} can be generalized
to the situation where $L$ is a certain type of torus ribbon knot in $S^2 \times S^1$.
In particular, it is very likely that using the approach above
and one can obtain  a rigorous continuum analogue of Theorem 5.7 in \cite{Ha9}.
\end{remark}

\section{Discussion \& Outlook}
\label{sec4}

\subsection{Open Questions}
\label{subsec4.1}

\begin{question} \label{quest0}
Are Conjectures \ref{conj2}--\ref{conj3}  above indeed true?
\end{question}

If the answer to Question \ref{quest0} is ``yes'', then one arrives naturally
at the following question:

\begin{question} \label{quest3}
Can  Assumption \ref{ass0} be dropped?
In other words: will the more or less straightforward\footnote{Recall that when Assumption \ref{ass0}
 is dropped we  need to modify Assumption \ref{ass1} (cf. Remark \ref{rm_Sec2.5_Ende} in Sec. \ref{subsec2.5})
 and some of the constructions \& definitions in Sec. \ref{subsec3.3}.
 Moreover, we need to give a (heuristic) derivation/justification for formula \eqref{eq_crucial}  in Sec. \ref{subsec2.6} also in the case of general ribbon links $L$.}
 generalizations of Conjectures \ref{conj2}--\ref{conj3}
to the case of  generic\footnote{in fact, we  expect that the class of ribbon links for which
our approach is applicable cannot be the class of general ribbon links $L = (R_1, R_2, \ldots, R_m)$.
All ``singular'' twists of the ribbons $R_i$, $ i \le m$, must probably be excluded. One sufficient condition
on $L$ which excludes such singular twists is that each $R^i_{\Sigma}$ is a local diffeomorphism.
Of course, crossings and self-crossing of the projected ribbons $R^i_{\Sigma}$, $i \le m$ (and certain ``regular'' twists) are nor excluded by this condition  }  ribbon links also be true?
\end{question}

Before one studies Question \ref{quest3} on a rigorous level
it is reasonable to consider this issue first on an informal level.

\smallskip

Apart from Questions \ref{quest0} and \ref{quest3}, which are obviously the main questions,
also the following two questions are of interest:

\begin{question} \label{quest1} Can  Assumption \ref{ass1} be dropped?
           If not, then is there a deeper reason  why we have to make such an
           assumption?
\end{question}

\begin{question} \label{quest2} Is it possible to find
regularized versions
  $ 1^{(n)}_{\cB_{reg}}$
  and  $\Det^{(n)}_{rig}$,  $n \in \bN$,
  of $ 1_{\cB_{reg}}: \cB \to \bR$
  and  $\Det_{rig}: \cB \to \bR$
  which are more natural than the ones given in Appendix \ref{appB}?
\end{question}

\subsection{Comparison with the simplicial approach}
\label{subsec4.2}

As mentioned in  Sec. \ref{sec1} there is an alternative approach
for making rigorous sense of the RHS of Eq. \eqref{eq2.48_Ha7a}, namely the  ``simplicial approach''
developed in  \cite{Ha7a,Ha7b}.
The simplicial approach is essentially elementary\footnote{with the exception of some general
results from Lie theory} and, as a result, rigorous proofs
can be obtained more easily than within the continuum approach
 of the present paper.
 Moreover, the simplicial approach is probably better suited
 for the kind of applications we have in mind (cf.  ``Problem 3'' in the Introduction in \cite{Ha7a}).\par

 In spite of this it is still important  to study and elaborate  the rigorous continuum approach
 of the present paper.
 The following list  should make clear:

\begin{itemize}

\item The continuum approach allows us to avoid  the transition to BF-theory, which is apparently
  necessary in the simplicial apprroach.


\item   In the simplicial approach in \cite{Ha7a,Ha7b} there is one issue which is not totally understood.
  In the rigorous realization of $\Det_{FP}(B)$ in  \cite{Ha7a,Ha7b} we need to include
  an $1/2$-exponent in order to get the correct result. At the moment we only have a rather vague justification
  for this inclusion (cf. Appendix D in \cite{Ha7b}).
  By contrast, in the rigorous continuum approach  of the present paper this issue does not play a role.

\item In view of Conjecture \ref{conj2} and  Conjecture  \ref{conj3}, for the special type of (ribbon) links $L$ fulfilling Assumption \ref{ass0}
   the approach of the present paper
   should lead to the same explicit expressions as the approach in \cite{Ha7a,Ha7b}.
   However, for general ribbon links $L$ this is most probably
   not the case. It may well turn out that for general ribbon links
   only the continuum approach will give us the correct expressions for the WLOs
   while the simplicial approach in its original form does not.

\item Finally, the rigorous version of continuum approach is closer to heuristic computations
 in \cite{Ha10}. So if the project in \cite{Ha10} can be carried out successfully,
 it will  almost certainly be clear that by adapting the approach of the present paper
 one can obtain a rigorous treatment within the framework of White Noise Analysis.

\end{itemize}

 \renewcommand{\thesection}{\Alph{section}}
\setcounter{section}{0}

\section{Appendix: Turaev's shadow invariant}
\label{appA}

Let us briefly recall the definition of
Turaev's shadow invariant in the  situation relevant for us, i.e. for the base manifold
$M=\Sigma \times S^1$ with  $\Sigma=S^2$.

\subsection{Lie theoretic notation}
\label{appA.1}

Let $G$, $T$, $\cG$, $\ct$,   $\langle \cdot, \cdot \rangle $, and $\ck$
 be as in Sec. \ref{sec2} above.
 Using the scalar product $\langle \cdot, \cdot \rangle$ we can make the identification $\ct \cong \ct^*$.
Let us now fix a Weyl chamber $\cC \subset \ct$
and  introduce the following notation:
\begin{itemize}

\item $\cR \subset \ct^*$: the set  of real roots associated to $(\cG,\ct)$

\item  $\cR_+ \subset \cR$:  the set of positive (real) roots
 corresponding to $\cC$

\item $\rho$: half sum of positive roots (``Weyl vector'')
\item $\theta$: unique long root in the Weyl chamber $\cC$.
\item  $\cg= 1 + \langle \theta,\rho \rangle$:  the dual Coxeter number of
 $\cG$.

\item $I \subset \ct$: the kernel of $\exp_{|\ct}:\ct \to T$.  We remark that from
the assumption that $G$ is simply-connected
 it follows that $I$ coincides with the lattice $\Gamma$ which is generated by the
set of real coroots associated to $(\cG,\ct)$.

\item  $\Lambda \subset \ct^* (\cong \ct)$:  the {\it real} weight lattice associated to $(\cG,\ct)$, i.e.
$\Lambda$ is the lattice which is dual to $\Gamma = I$.

\item $\Lambda_+ \subset \Lambda$:  the set of  dominant  weights corresponding to $\cC$,
i.e. $\Lambda_+ := \bar{\cC} \cap \Lambda$

\item  $\Lambda^k_+ \subset \Lambda$, $k \in \bN$: the subset of $\Lambda_+$ given by
$\Lambda^k_+ :=   \{ \l \in \Lambda_+  \mid  \langle \l ,\th \rangle \leq k - \cg \}$
(the ``set of dominant weights which are integrable at level $l := k - \cg$'').

\item $\cW \subset \GL(\ct)$: the Weyl group of the pair $(\cG,\ct)$

\item $\cW_{\aff} \subset \Aff(\ct)$: the ``affine Weyl group of  $(\cG,\ct)$'',
i.e. the subgroup of $\Aff(\ct)$ generated by $\cW$ and the set of translations $\{ \tau_x \mid x \in \Gamma\}$
where $\tau_x: \ct \ni b \mapsto b + x \in \ct$.

\item $\cW_{k} \subset \Aff(\ct)$, $k \in \bN$:  the subgroup of $\Aff(\ct)$
given by $\{ \psi_k \circ \sigma \circ \psi_k^{-1} \mid \sigma \in \cW_{\aff} \}$
where $\psi_k : \ct \in b \mapsto b \cdot k - \rho \in \ct$
(the ``quantum Weyl group corresponding to the  level $l := k - \cg$'')

\end{itemize}

The following formulas are used in Sec. \ref{subsec2.5}  above and Appendix \ref{appB} below.
For $b \in \ct$ we have
\begin{equation} \label{eq_appA2}
b \in \ct_{reg} \quad \Leftrightarrow \quad [ \forall \alpha \in \cR_+: \ \alpha(b) \notin \bZ]
\end{equation}
and
\begin{multline} \label{eq_appA1}
\det(\id_{\ck}-\exp(\ad(b))_{|\ck}) = \prod_{\alpha \in \cR} (1- e^{2\pi i \alpha(b)})\\
= \prod_{\alpha \in \cR+} (1- e^{2\pi i \alpha(b)}) (1- e^{- 2\pi i \alpha(b)})
=  \prod_{\alpha \in \cR+}  4 \sin^2(\pi  \alpha(b))
\end{multline}

\subsection{The shadow invariant}
\label{appA.2}

Let $L= (l_1, l_2, \ldots, l_m)$, $m \in \bN$,  be a framed  link
in  $M= \Sigma \times S^1$.
For simplicity we will assume that each $l_i$, $i \le m$ is equipped with a ``horizontal'' framing\footnote{here
we use the terminology of Remark 4.5  in \cite{Ha7a}. We remark that in the special case  when
   $L$ is the framed link which is induced by a ribbon link $L = L_{ribb}$  fulfilling Assumption \ref{ass0} in  Sec. \ref{subsec2.4} then each $l_i$ will automatically be equipped with a  horizontal framing}.
Let $V(L)$ denote the set of points $p \in \Sigma$ where the loops
 $l^i_{\Sigma}$, $i \le m$, cross themselves or each other (the ``crossing points'')
 and $E(L)$ the set of
curves in $\Sigma$ into which the loops $l^1_{\Sigma}, l^2_{\Sigma},
\ldots, l^m_{\Sigma}$ are decomposed when being ``cut''  in the
points of $V(L)$.
 We assume that there are only finitely many
 connected components $Y_0, Y_1, Y_2, \ldots, Y_{m'}$, $m' \in \bN$ (``faces'')
 of  $\Sigma \backslash ( \bigcup_i \arc(l^i_{\Sigma}))$ and set
 $$F(L):= \{ Y_0, Y_1, Y_2, \ldots, Y_{m'} \}.$$
As explained in \cite{Tu2} one can associate in a natural way a
 half integer $\gleam(Y)  \in \tfrac{1}{2} \bZ$, called ``gleam'' of $Y$,
to each face $Y \in F(L)$. In the special case  where the link $L$ is a framed link
which is induced by a ribbon link $L = L_{ribb}$  fulfilling Assumption \ref{ass0} in  Sec. \ref{subsec2.4}
  we have the explicit formula
\begin{equation} \label{eqA.1}
\gleam(Y) = \sum_{i \text{ with } \arc(l^i_{\Sigma}) \subset
\partial Y}  \wind(l^i_{S^1}) \cdot \sgn(Y;l^i_{\Sigma}) \in \bZ
\end{equation}
where $ \wind(l^i_{S^1})$ is the winding number of the loop
 $l^i_{S^1}$ and where $ \sgn(Y;l^i_{\Sigma})$
 is given by
 \begin{equation} \label{eqA.2}
  \sgn(Y; l^{i}_{\Sigma}):=
\begin{cases} 1 & \text{ if  $Y \subset Z^+_i$ }\\
-1 & \text{ if  $Y \subset Z^-_i$ }\\
\end{cases}
\end{equation} Here  $Z^{+}_i$ (resp. $Z^{-}_i$) is the unique
connected component $Z$ of $\Sigma \backslash \arc(l^i_{\Sigma})$
such that $l^i_{\Sigma}$ runs around $Z$ in the ``positive'' (resp.
``negative'') direction. \par

Assume that each loop $l_i$ in the link $L$  is equipped with a ``color'' $\rho_i$,
i.e. a finite-dimensional complex representation of $G$.
By   $\gamma_i \in \Lambda_+$ we denote the highest weight of
 $\rho_i$ and set     $\gamma(e):= \gamma_i$
 for each $e \in E(L)$  where $i \le n$ denotes the unique
    index such that  $\arc(e) \subset \arc(l_i)$.
Finally, let  $col(L)$ be the set of all mappings $\vf: \{Y_0,
Y_1, Y_2, \ldots, Y_{m'}\} \to \Lambda^k_+$ (``area
colorings'').\par

We can now define the ``shadow invariant'' $|L|$ of the
(colored and ``horizontally framed'')   link $L$
associated to the pair $(\cG,k)$  by
 \begin{equation} \label{eqA.4}
|L|:= \sum_{\vf\in col(L)}
|L|_1^{\vf}\,|L|_2^\vf\,|L|_3^\vf\,|L|_4^\vf~
\end{equation}
  with
  \begin{subequations} \label{eqA.5}
\begin{align} |L|_1^\vf&=\prod_{Y \in F(L)} \dim(\vf(Y))^{\chi(Y)}\\
|L|_2^\vf&= \prod_{Y \in F(L)}   \exp(\tfrac{\pi i}{{k}} \langle \vf(Y),\vf(Y) +2\r\rangle)^{\gleam(Y)}\\
\label{eq_XL3}  |L|_3^\vf&= \prod_{e \in E_*(L)} N_{\gamma(e)
\varphi(Y^+_e)}^{\varphi(Y^-_e)} \\
|L|_4^\vf&= \bigl( \prod_{e \in E(L) \backslash E_*(L)} S(e,\vf) \bigr) \times \bigl( \prod_{x \in V(L)} T(x,\vf) \bigr)
  \end{align}
 \end{subequations}
 Here $Y^+_e$ (resp. $Y^-_e$) denotes the  unique face $Y$
  such that $\arc(e) \subset \partial Y$ and,
  additionally,  the orientation on $\arc(e)$ described above
coincides with (resp. is opposite to) the orientation which is
obtained by restricting the orientation on $\partial Y$ to $e$.
Moreover,    we have set
 (for $\lambda, \mu, \nu \in \Lambda^k_+$)
  \begin{equation}  \label{eqA.6}
\dim(\lambda) := \prod_{\a \in \cR_+}{\sin{\pi \langle \l+\r,\a
\rangle \over {k} }\over\sin{\pi \langle \r,\a \rangle \over
{k} }}
 \end{equation}
    \begin{equation} \label{eqA.7}
 N_{ \mu \nu}^{\lambda} := \sum_{\tau \in \cW_{{k}}}
  \sgn(\tau) m_{\mu}(\nu-\tau(\lambda))
 \end{equation}
where $m_{\mu}(\beta)$ is the multiplicity of the weight $\beta$
in the  unique (up to equivalence) irreducible
representation $\rho_{\mu}$ with  highest weight $\mu$
and   $\cW_{{k}}$ is as above.
$E_*(L)$ is a suitable subset of $E(L)$ (cf. the notion of ``circle-1-strata''
in  Chap. X, Sec. 1.2 in \cite{Turaev}). \par

The  explicit expression for the factors  $T(x,\vf)$ appearing in $|L|_4^\vf$  above
involves the so-called  ``quantum 6j-symbols'' (cf. Chap. X, Sec. 1.2 in \cite{Turaev}) associated to
the quantum group $U_q(\cG_{\bC})$
 where $q$ is the root of unity
\begin{equation} \label{eq_rootunity} q:= \exp( \tfrac{2 \pi i}{k})
\end{equation}
We omit the explicit formulae for $T(x,\vf)$ and $S(e,\vf)$ since they irrelevant for our purposes.
Indeed, if $L$ is the framed link
which is induced by a ribbon link $L = L_{ribb}$  fulfilling Assumption \ref{ass0} in  Sec. \ref{subsec2.4}
  the set $V(L)$ is empty and
the set $E_*(L)$ coincides with $E(L)$, so  Eq. \eqref{eqA.4}  then reduces to
 \begin{equation} \label{eqA.8}
|L| = \sum_{\vf\in col(L)} \biggl( \prod_{i=1}^m N_{\gamma(l_i)
\varphi(Y^+_{i})}^{\varphi(Y^-_{i})} \biggr)  \biggl(  \prod_{Y \in F(L)}
\dim(\vf(Y))^{\chi(Y)}
  \exp(\tfrac{\pi i}{{k}} \langle \vf(Y),\vf(Y) +2\r \rangle )^{\gleam(Y)} \biggr)
  \end{equation}
   where  we have set $Y^{\pm}_{i}:= Y^{\pm}_{l^i_{\Sigma}}$.

\section{Appendix: Construction of $ 1^{(n)}_{\cB_{reg}}$ and $\Det^{(n)}_{rig}$}
\label{appB}

We will now describe the  regularized versions
$ 1^{(n)}_{\cB_{reg}}: \cB \to \bR$
  and  $\Det^{(n)}_{rig}: \cB \to \bR$,  $n \in \bN$,
  of the functions $1_{\cB_{reg}}$ and $\Det_{rig}$
  which we used in Sec. \ref{subsec3.4} above.

\begin{itemize}
\item  Let $\triangle_0$ be a fixed finite triangulation  of $\Sigma$
which is ``compatible'' with $L$ in the sense that $\arc(L_{\Sigma})$ is contained in the 1-skeleton of $\triangle_0$.
 For each $n \in \bN$ let $\triangle_n$ be the barycentric sub division of $\triangle_{n-1}$.
 We denote by $\faces_2(\triangle_n)$  the set of 2-faces of $\triangle_n$.
  For $B \in \cB$ and $F \in \faces_2(\triangle_n)$ let $B(F)$  be the ``mean value'' of $B$ on $F$,
  i.e.
 \begin{equation} B(F) := \frac{\int_F B d\mu_{\mathbf g}}{\int_F 1 d\mu_{\mathbf g}} \in \ct
 \end{equation}

\item We approximate the indicator function
 $1_{\ct_{reg}}:\ct \to [0,1]$
 by a suitable sequence of trigonometric polynomials\footnote{in order to see that such a sequence
 always exist we first choose, for each fixed $n \in \bN$, a smooth $1$-periodic function
 $\psi^{(n)}: \bR \to \bR$ such that $\psi^{(n)}(x) = 0$ for all $x \in \bZ$ and
 $\psi^{(n)}(x) = 1$ for all $x$ outside the $\tfrac{C}{n}$-neighborhood of $\bZ \subset \bR$ for some fixed $C > 0$.
   Since  $\psi^{(n)}$ is a smooth periodic function  its  Fourier series  converges uniformly.
 Accordingly, for every fixed $\eps = \eps_n >0$ we can find a (1-periodic) trigonometric polynomial $p^{(n)}$
 such that $\| \psi^{(n)} -  p^{(n)}\|_{\infty} < \eps$ and, consequently,
 $\| \psi^{(n)} -  \bar{p}^{(n)}\|_{\infty} < 2\eps$ where $\bar{p}^{(n)} := p^{(n)} - p^{(n)}(0)$.
 (Clearly, $\bar{p}^{(n)}(x) = 0$ for all $x \in \bZ$.)
  In view of the identity
 $1_{\ct_{reg}}(b) = \prod_{\alpha \in \cR_+} 1_{\bR \backslash \bZ}(\alpha(b))$, cf.
  Eq. \eqref{eq_appA2} above, we now define $1_{\ct_{reg}}^{(n)}$ by
   $1_{\ct_{reg}}^{(n)}(b): =   \prod_{\alpha \in \cR_+} \bar{p}^{(n)}(\alpha(b)) $ for all $b \in \ct$.
   Clearly, if $C >0$ was chosen small enough  and for every $n \in \bN$ the number $\eps = \eps_n$ was chosen small enough  we obtain a family $(1_{\ct_{reg}}^{(n)})_{n \in \bN}$ with the desired properties}
  $(1_{\ct_{reg}}^{(n)})_{n \in \bN}$ such that
  for all $n \in \bN$ we have
 \begin{align*}
  1_{\ct_{reg}}^{(n)}(b) & = 0 \quad \text{for all $b \in  \ct_{sing}:= \ct \backslash \ct_{reg}$,  and } \\
  | 1_{\ct_{reg}}^{(n)}(b) -   1| & \le 1/N_n^2 \quad
  \text{for all $b \in \ct$  outside the $1/n$-neighborhood of $\ct_{sing}$}
 \end{align*}
 where we have set  $N_n := \# \faces_2(\triangle_n)$. \par

For each $n \in \bN$ we  then introduce  $ 1^{(n)}_{\cB_{reg}}: \cB \to \bR$ by
$$  1^{(n)}_{\cB_{reg}}(B)  = \prod_{F \in \faces_2(\triangle_n)} 1_{\ct_{reg}}^{(n)}\bigl(B(F)\bigr)  \quad \forall B \in \cB $$

\item  Recall from  Sec. \ref{subsec2.5} that
 for  $B \in \cB_{reg}$ we have
 $$\Det_{rig,\alpha}(B) =
 \exp\biggl( \int_{\Sigma} \bigl[ \log\bigl(2 \sin(\pi  \alpha(B(\sigma)))\bigr)  \tfrac{1}{4 \pi}   R_{\mathbf g} (\sigma) \bigr] d\mu_{\mathbf g}(\sigma)\biggr)$$
 where $\log: \bR \backslash \{0\} \to \bC$ is the restriction to  $\bR \backslash \{0\}$
 of the principal branch of the complex logarithm, i.e. is given by
 $$\log(x) = \ln(|x|) + \pi i H(-x) \quad \forall x \in \bR \backslash \{0\}$$
 where $H(x) = (1 + \sgn(x))/2$ is the Heaviside function.

  For $B \in \cB$ and  $n \in \bN$ we now define the ``step function''
 $B^{(n)}: \Sigma \to \ct$ by
 \begin{equation} \label{eq_def_Bn}
 B^{(n)} = \sum_{F \in \faces_2(\triangle_n)} B(F) \cdot 1_{F}
  \end{equation}
  where $B(F)$ is as above.  Moreover, for fixed $n \in \bN$ we now replace $\exp(x)$, $x \in \bR$,
 by the $n$th Taylor polynomial
$\exp^{(n)}(x)= \sum_{k=0}^n \tfrac{x^k}{k!}$
and  $\log(x)$ by $\log^{(n)}(x)$ where
$(\log^{(n)})_{n \in \bN}$ is any fixed sequence of polynomial functions $\log^{(n)}(x): I \to \bC$
which converges uniformly to $\log$
on every compact subinterval of $I:= [-2,2] \backslash \{0\}$. \par

After these preparations we  set for each $B \in \cB$.
\begin{subequations} \label{eq_Detn_Def}
 \begin{equation} \Det^{(n)}_{rig,\alpha}(B) :=
 \exp^{(n)}\biggl( \int_{\Sigma} \bigl[ \log^{(n)}(2 \sin(\pi  \alpha(B^{(n)}(\sigma))))  \tfrac{1}{4 \pi}   R_{\mathbf g} (\sigma) \bigr] d\mu_{\mathbf g}(\sigma)\biggr)
 \end{equation}

and define $\Det^{(n)}_{rig}: \cB \to \bR$, $n \in \bN$,  by
 \begin{equation} \Det^{(n)}_{rig}(B)    := \prod_{\alpha \in \cR_+}  \Det^{(n)}_{rig,\alpha}(B) \quad \forall B \in \cB
 \end{equation}
\end{subequations}
\end{itemize}

\begin{remark} \label{rm_appB} \rm
\begin{enumerate}

\item  After carrying out the $\eps \to 0$ and the $s \to 0$-limits on the RHS of Eq. \eqref{eq_limit1}
step functions of the type  $B= \sum_{i=1}^r b_i 1_{Y_i}$ appear in the calculations, cf. the paragraph after
 Eq. \eqref{eq_step_function} in Sec. \ref{subsec2.5}.
It is useful (and straightforward) to generalize the definitions of
 $ 1^{(n)}_{\cB_{reg}}(B)$ and $\Det^{(n)}_{rig}(B)$
 to such functions $B$.

\item In order to justify that $(1^{(n)}_{\cB_{reg}})_{n \in \bN}$
 is indeed a regularization of $1_{\cB_{reg}}$
 observe that if $B \in \cB_{reg}$ then
 $$ \lim_{n \to \infty} 1^{(n)}_{\cB_{reg}}(B) = 1$$
 On the other hand, we do not necessarily have $\lim_{n \to \infty} 1^{(n)}_{\cB_{reg}}(B) = 0$
 if $B \notin \cB_{reg}$. However,
   if $B$ is a step function of the type $B= \sum_{i=1}^r b_i 1_{Y_i}$ as in Eq. \eqref{eq_step_function} in Sec. \ref{subsec2.5} (which is the only case which will play a role in our computations after
   the  $\eps \to 0$ and the $s \to 0$-limits on the RHS of Eq. \eqref{eq_limit1} have been taken)
   then we do have\footnote{Here we use the condition mentioned above that the  triangulation $\triangle_0$   of $\Sigma$  and therefore also all barycentric subdivisions $\triangle_n$ of $\triangle_0$
 are compatible with $L$ in the sense above}
 $$ \lim_{n \to \infty} 1^{(n)}_{\cB_{reg}}(B) =
 \begin{cases}  1 & \text{ if $B$ takes only values in $\ct_{reg}$}\\
  0 & \text{ if $B$ takes at least one  value in $\ct_{sing}$}
 \end{cases}$$
where $1^{(n)}_{\cB_{reg}}(B)$ is the aforementioned generalization.

\item In order to justify that $(\Det^{(n)}_{rig})_{n \in \bN}$ is  a regularization
of $\Det_{rig}(B)$ we will now verify that
for  all  $B \in \cB_{reg}$ we have indeed
\begin{equation} \label{eq_appB_last} \lim_{n \to \infty} \Det^{(n)}_{rig}(B) = \Det_{rig}(B)
\end{equation}
 Observe  first that since $B \in \cB_{reg}$
 we have    $B(\sigma) \in \ct_{reg}$  and therefore (cf. Eq. \eqref{eq_appA2})
  $ \sin(\pi  \alpha(B(\sigma)) \neq 0$ for all $\sigma \in \Sigma$.
 Now let $\cS_n$ be the 1-skeleton of $\triangle_n$
 and let $\cN = \bigcup_{n \in \bN} \cS_n$.
  Then  for all $\sigma \in \Sigma$ which do not lie in $\cN$
we have $\lim_{n \to \infty} B^{(n)}(\sigma) = B(\sigma)$ and therefore  also
$  \sin(\pi  \alpha(B^{(n)}(\sigma)) \neq 0$
if $n \in \bN$ is sufficiently large. According to the choice of $\log^{(n)}$
we therefore obtain
\begin{equation} \label{eq_appB_last2} \lim_{n \to \infty} \log^{(n)}(2 \sin(\pi  \alpha(B^{(n)}(\sigma)))
=  \log( 2 \sin(\pi  \alpha(B(\sigma)))
\end{equation}
Since $\cN$ is a $\mu_{\mathbf g}$-zero set
Eq. \eqref{eq_appB_last2} holds for $\mu_{\mathbf g}$-almost all $\sigma \in \Sigma$.
From  Eqs. \eqref{eq_Detn_Def} it  now easily follows that
Eq. \eqref{eq_appB_last} is indeed fulfilled.  \par

Finally, observe that for step functions $B= \sum_{i=1}^r b_i 1_{Y_i}$
of the type mentioned above
which satisfy $1^{(n)}_{\cB_{reg}}(B) \neq 0$ for sufficiently large $n$
we have\footnote{Here we use again that every  $\triangle_n$ is compatible with $L$}
 again Eq. \eqref{eq_appB_last}
(and, in fact, even  $\Det^{(n)}_{rig}(B) = \Det_{rig}(B)$ for sufficiently large $n$).
\end{enumerate}
\end{remark}

I want to emphasize once more that the regularized versions
$ 1^{(n)}_{\cB_{reg}}: \cB \to \bR$
  and  $\Det^{(n)}_{rig}: \cB \to \bR$
  of the functions $1_{\cB_{reg}}$ and $\Det_{rig}$
  introduced above are probably not the best ones.
   It would be desirable  to find a more elegant and more natural regularization
   (cf. Question \ref{quest2} in Sec. \ref{sec4} above).
In particular, one should try to avoid the use of triangulations, which is clearly not in the spirit
of a continuum approach.


\begin{thebibliography}{10}

\bibitem{ASen}
S.~Albeverio and A.N. Sengupta.
\newblock {A Mathematical Construction of the Non-Abelian Chern-Simons
  Functional Integral}.
\newblock {\em Commun. Math. Phys.}, 186:563--579, 1997.


\bibitem{BlTh1}
M.~Blau and G.~Thompson.
\newblock {Derivation of the Verlinde Formula from Chern-Simons Theory and the
  G/G model}.
\newblock {\em Nucl. Phys.}, B408(1):345--390, 1993.

\bibitem{BlTh2}
M.~Blau and G.~Thompson.
\newblock {Lectures on 2d Gauge Theories: Topological Aspects and Path Integral
  Techniques}.
\newblock In E.~Gava et~al., editor, {\em Proceedings of the 1993 Trieste
  Summer School on High Energy Physics and Cosmology}, pages 175--244. World
  Scientific, Singapore, 1994.

\bibitem{BlTh3}
M.~Blau and G.~Thompson.
\newblock {On Diagonalization in $Map(M,G)$}.
\newblock {\em Commun. Math. Phys.}, 171:639--660, 1995.


\bibitem{dFPS}
M.~de~Faria, J.~Potthoff, and L.~Streit.
\newblock The {F}eynman integrand as a {H}ida distribution.
\newblock {\em J. Math. Phys.}, 32(8):2123--2127, 1991.



\bibitem{HaHa}
S.~de~Haro and A.~Hahn.
\newblock {Chern-Simons theory and the quantum Racah formula}.
\newblock {\it Rev. Math. Phys.}, 25:1350004, 41 pp., 2013
[arXiv:math-ph/0611084]



\bibitem{Ha2}
A.~Hahn.
\newblock {The Wilson loop observables of Chern-Simons theory on ${\mathbb
  R}^3$ in axial gauge}.
\newblock {\em {Commun. Math. Phys.}}, 248(3):467--499, 2004.

\bibitem{Ha3b}
A.~Hahn.
\newblock {Chern-Simons models on $S^2 \times S^1$, torus gauge fixing, and
  link invariants I}.
\newblock {\em {J. Geom. Phys.}}, 53(3):275--314, 2005.


\bibitem{Ha3c} A.~Hahn.
\newblock {Chern-Simons models on $S^2 \times S^1$, torus gauge fixing, and
  link invariants II}.
\newblock {\em {J. Geom. Phys.}}, 58:1124--1136, 2008.

\bibitem{Ha4}
A.~Hahn.
\newblock {An analytic Approach to Turaev's Shadow Invariant}.
\newblock {\em J. Knot Th. Ram.}, 17(11): 1327--1385, 2008
[see arXiv:math-ph/0507040v7 (2011) for the most recent version]



\bibitem{Ha6}
A.~Hahn.
\newblock {White noise analysis in the theory of three-manifold quantum
  invariants}.
\newblock In A.N. Sengupta and P.~Sundar, editors, {\it Infinite Dimensional
  Stochastic Analysis}, volume XXII of {\it Quantum Probability and White Noise
  Analysis}, pages 201--225. World Scientific, 2008.

\bibitem{Ha7a}
A.~Hahn.
\newblock {From simplicial Chern-Simons theory to the shadow invariant I},
 J. Math. Phys. 56: 032301, 52 pp., 2015  [arXiv:1206.0439]

\bibitem{Ha7b}
A.~Hahn.
\newblock {From simplicial Chern-Simons theory to the shadow invariant II},
  J. Math. Phys.  56: 032302, 46 pp., 2015   [arXiv:1206.0441].


\bibitem{Ha9}
A.~Hahn. Torus Knots and the Chern-Simons path integral: a rigorous treatment. Preprint, 2015
[arXiv:1508.03804].


\bibitem{Ha10}
A.~Hahn. $R$-matrices and the Chern-Simons path integral. In preparation.


\bibitem{HKPS}
T.~Hida, H.-H. Kuo, J.~Potthoff, and L.~Streit.
\newblock {\em {White Noise. An infinite dimensional Calculus}}.
\newblock Dordrecht: Kluwer, 1993.

\bibitem{KLPSW}
 Y. Kondratiev, P. Leukert, J. Potthoff, L. Streit, W. Westerkamp.
\newblock { {Generalized Functionals in Gaussian Spaces -- the Characterization Theorem Revisited}}.
 \newblock {\em  J. Funct. Anal.}, 141(2),  301--318, 1996

\bibitem{KPS} H.-H. Kuo, J.~Potthoff, and L.~Streit.
\newblock {A characterization of white noise test functionals}.
 \newblock {\em  Nagoya Math. J.}, 121,  185--194, 1991

\bibitem{Kup}
G.~Kuperberg.
\newblock {Quantum invariants of knots and 3-manifolds (book review)}.
\newblock {\em Bull. Amer. Math. Soc.}, 33(1):107--110, 1996.


\bibitem{LS}
P.~Leukert and J.~Sch{\"a}fer.
\newblock {A Rigorous Construction of Abelian Chern-Simons Path Integrals using
  White Noise Analysis}.
\newblock {\em Rev. Math. Phys.}, 8(3):445--456, 1996.



\bibitem{McSin}
H. P.~ McKean and I.~M. Singer.
\newblock {Curvature and the eigenvalues of the Laplacian}.
\newblock {\em J.~Differential Geometry}, 1(1):43--69, 1967.


\bibitem{Pat}
V. K.~ Patodi.
\newblock {Curvature and the eigenforms of the Laplace Operator}.
\newblock {\em J.~Differential Geometry}, 5(1):233--249, 1971.


\bibitem{ReTu2}
N.Y. Reshetikhin and V.G. Turaev.
\newblock {Ribbon graphs and their invariants derived from quantum groups.}
\newblock {\em Commun. Math. Phys.}, 127:1--26, 1990.

\bibitem{ReTu1}
N.Y. Reshetikhin and V.G. Turaev.
\newblock {Invariants of three manifolds via link polynomials and quantum
  groups.}
\newblock {\em Invent. Math.}, 103:547--597, 1991.

\bibitem{Sawin99}
S.~Sawin.
\newblock {Jones-Witten invariants for non-simply connected Lie groups and the geometry of the Weyl alcove}.
arXiv: math.QA/9905010, 1999.


\bibitem{HiSt}
L.~Streit and T.~Hida.
\newblock Generalized {B}rownian functionals and the {F}eynman integral.
\newblock {\em Stochastic Process. Appl.}, 16(1):55--69, 1984.


\bibitem{Tu2} V.~G.~Turaev, ``Shadow Links and Face Models of Statistical Mechanics'', J.~Differential
Geometry 36:35-74, 1992

\bibitem{Turaev} V.~G.~Turaev, ``Quantum Invariants of Knots and 3-Manifolds'',  De Gruyter, 1994


\bibitem{Wi}
E.~Witten.
\newblock {Quantum Field Theory and the Jones Polynomial}.
\newblock {\em Commun. Math. Phys.}, 121:351--399, 1989.


\end{thebibliography}
\end{document}